\documentclass[12pt]{article}

\usepackage{color}
\definecolor{darkblue}{rgb}{0.1,0.1,.7}
\usepackage[colorlinks, linkcolor=darkblue, citecolor=darkblue, urlcolor=darkblue, linktocpage]{hyperref}
\usepackage[]{amsmath}
\usepackage[]{graphicx}
\usepackage[]{latexsym}
\usepackage{slashed,graphicx,color,amsmath,amssymb}
\usepackage[mathscr]{eucal}
\usepackage{mathrsfs}
\usepackage{geometry}
\usepackage[margin=10pt,font=small,labelfont=bf]{caption}
\usepackage{amscd}
\usepackage{bm}
\usepackage{xcolor}
\usepackage{upgreek}
\usepackage[square, comma, sort&compress,numbers]{natbib}
\usepackage[all,cmtip]{xy}
\usepackage[color=cyan!30!white,linecolor=red,textsize=footnotesize]{todonotes}
\numberwithin{equation}{section}
 \makeatletter
 \g@addto@macro\bfseries{\boldmath}
 \makeatother
\usepackage{amsmath}
\begin{document}
\vspace*{-.6in} \thispagestyle{empty}
\vspace{.2in} {
\begin{center}
{\bf Root-$T\overline{T}$ deformed CFT partition functions at large central charge}
\end{center}}
\vspace{.2in}
\begin{center}
\vspace{.1in}
\renewcommand{\thefootnote}{\fnsymbol{footnote}}
\begin{center}
{Miao He\footnotemark[1]\footnotetext[1]{E-mail: miaohe18@gmail.com}}
\vspace{.2in}\\
\textit{School of Physics and Electronic Information, Anhui Normal University \\ Wuhu 241000, China}
\end{center}
\end{center}
\vspace{.3in}

\begin{abstract}
In this work, we investigate the partition function of 2d CFT under root-$T\bar{T}$ deformation. We demonstrate that the deformed partition function satisfies a flow equation. At large central charge sector, the deformed partition function reduces to a redefinition of the modular parameters, which preserves modular invariance under the deformed parameters. We then derive a Cardy-like formula for the asymptotic density of states using modular bootstrap trick. In the context of AdS/CFT, it was proposed the root-$T\bar{T}$ deformed CFT corresponds to the AdS$_3$ with certain deformed boundary condition. We show the deformed BTZ black hole is a quotient of hyperbolic space. In terms of Chern-Simons formulation, we compute the root-$T\bar{T}$ deformed BTZ black hole entropy and find that it obeys a Cardy-like formula, which is consistent with the modular bootstrap result. Furthermore, employing the Wilson spool technique, we compute the one-loop partition functions for the root-$T\bar{T}$ deformed AdS$_3$ geometry. Our results reveal an exact agreement between one-loop gravitational partition function and the large $c$ expansion of root-$T\bar{T}$ deformed CFT partition function.
\end{abstract}

\vskip 1cm \hspace{0.7cm}

\newpage

\setcounter{page}{1}
\begingroup
\hypersetup{linkcolor=black}
\tableofcontents
\endgroup
\renewcommand{\thefootnote}{\arabic{footnote}}
\section{Introduction}
\label{sec:1}
The $T\bar{T}$ deformations have been proposed and developed for about ten years~\cite{Smirnov:2016lqw,Cavaglia:2016oda}. Its importance stems from two key features: preservation of integrability and connections to holography. The solvability of this deformation manifests through several exact results, such as the deformed Lagrangian satisfies a flow equation~\cite{Bonelli:2018kik}, the spectrum obeys an inviscid Burgers equation, and the partition function follows a differential equation~\cite{Datta:2018thy,Dubovsky:2018bmo,Aharony:2018bad}. Moreover, it modifies S-matrices of integrable field theory by multiplying the universal CDD factor. Despite these exact results, $T\bar{T}$ deformation introduces novel UV behaviours characterized by non-locality, making correlation function and entanglement entropy computations particularly challenging. The perturbative approaches have been developed in~\cite{Guica:2019vnb,He:2019vzf,He:2020qcs,He:2020udl,He:2020cxp,He:2023kgq}. It is shown that it is more convenient to study the correlation functions in momentum space~\cite{Aharony:2018vux,Guica:2021fkv,Guica:2022gts,Chakraborty:2023wel}. Following Cardy's pioneering work on correlation functions~\cite{Cardy:2019qao}, recent advances employ the holographic worldsheet techniques~\cite{Cui:2023jrb} and 2d gravity descriptions~\cite{Aharony:2023dod,Barel:2024dgv,Hirano:2024eab}. This kind of deformation also attracts a lot of interests in holography~\cite{Chakraborty:2018kpr,Asrat:2017tzd,Giveon:2017nie,McGough:2016lol,Kraus:2018xrn,Cardy:2018sdv,Callebaut:2019omt,Tolley:2019nmm,Guica:2019nzm}. On the holographic aspects, the $T\bar{T}$-deformed 2d CFT corresponds to the AdS$_3$ gravity at finite radial cut-off~\cite{McGough:2016lol}. The cut-off radius is related to the deformation parameter via $\mu\sim 1/{r_c^2}$. The deformation is turned off as the cut-off radius goes to infinity, which recovers the standard AdS/CFT correspondence. Another alternative holographic description is imposing a mixed boundary condition at the asymptotic boundary of AdS$_3$~\cite{Guica:2019nzm}. The AdS$_3$ solutions with mixed boundary condition can be obtained by a field-dependent coordinate transformation~\cite{Guica:2019nzm}. The dynamical change of coordinate was also found from the field theory~\cite{Conti:2018tca,Conti:2019dxg}, as well as the 2D JT-like gravity~\cite{Dubovsky:2018bmo,Caputa:2020lpa}. For the mixed boundary condition holographic description, the $T\bar{T}$-deformed spectrum, Lagrangian and asymptotic symmetries can be reproduced in terms of Chern-Simons formulation of AdS$_3$~\cite{Ouyang:2020rpq,Llabres:2019jtx,He:2020hhm,He:2021bhj,Ebert:2022ehb}. The other studies about the holographic aspects of $T\bar{T}$ deformation can be found in~\cite{Chen:2018eqk,Caputa:2020lpa,Kraus:2021cwf,Ebert:2022cle,Kraus:2022mnu,Hirano:2020nwq,Jeong:2019ylz,Jeong:2022jmp,He:2023xnb,Chen:2023eic,Tian:2023fgf,Bhattacharyya:2023gvg,Poddar:2023ljf,Khoeini-Moghaddam:2020ymm, Apolo:2023vnm,Apolo:2023ckr, Chang:2024voo,Basu:2024enr, Basu:2024xjq, Basu:2025exh, Basu:2025fsf}. See~\cite{Jiang:2019epa,He:2025ppz} for review.
\par
Recently a new type of marginal deformation was proposed, which is called root-$T\bar{T}$ deformation~\cite{Ferko:2022cix,Babaei-Aghbolagh:2022uij,Babaei-Aghbolagh:2022leo,Babaei-Aghbolagh:2024uqp}. The 2D root-$T\bar{T}$ deformation can be obtained from the 4D root-$T\bar{T}$-like deformation, which deforms the Maxwell theory to the Modified Maxwell theory, through the dimension reduction~\cite{Conti:2022egv}. Although much less is known about the root-$T\bar{T}$ operator on quantum level, the root-$T\bar{T}$ deformation have been understand well classically. It preserves the classical integrability~\cite{Borsato:2022tmu} and conformal symmetry, since this is a marginal deformation. It is shown that this deformation might lead to an interesting class of models, including 2D field theory, 4D ModMax theory, and other higher dimension theory~\cite{Bandos:2020jsw, Hou:2022csf,Morone:2024ffm,Babaei-Aghbolagh:2025lko,Babaei-Aghbolagh:2025cni,Borsato:2022tmu,Ferko:2023ozb,Ferko:2024zth,Ebert:2024zwv,Bielli:2024khq,Ferko:2024yua}. Analogous to $T\bar{T}$, the root-$T\bar{T}$ deformation admits a 2D gravity description as well~\cite{Babaei-Aghbolagh:2024hti,Tsolakidis:2024wut}. The root-$T\bar{T}$ formation was also considered to study the BMS$_3$ symmetry of ultra/non-relativistic limits of CFT$_2$~\cite{Rodriguez:2021tcz,Tempo:2022ndz,Bagchi:2022nvj, Bagchi:2024unl}. However, the root-$T\bar{T}$ deformation becomes subtle on the quantum level, because there is no clear definition of the root-$T\bar{T}$ operator so far. We do not even know whether the root-$T\bar{T}$ operator is local operator. One of the possible definitions of the root-$T\bar{T}$ operator was proposed~\cite{Hadasz:2024pew}, in which the correlation functions were calculated perturbatively. 
\par
Another interesting feature is the holographic description of the root-$T\bar{T}$ deformation. It was proposed that the root-$T\bar{T}$ deformed CFT is dual to AdS$_3$ with a certain modified boundary condition~\cite{Ebert:2023tih}. The deformed bulk geometry can be obtained through a field-dependent coordinate transformation. The spectrum and boundary dynamics of this correspondence have been explicitly verified. From the holographic side, the large $c$ behaviour of the root-$T\bar{T}$ deformed holographic CFT was studied~\cite{Tian:2024vln}. The classical gravitational on-shell action is shown to be modular invariant. While on the field theory side, the calculation is still lacking. 
\par
In this paper, we shall study the root-$T\bar{T}$ deformation by analysing the partition function in the large $c$ limit, which allows for consistency check in holography. On the field theory side, we start from the deformed spectrum which satisfies a flow equation. By inserting the deformed spectrum into the torus partition function and expanding in powers of the deformation parameter, we derive the perturbative expansion for the root-$T\bar{T}$ deformed partition function. Our analysis reveals a recurrence relation, which leads to the flow equation for the deformed partition function. In addition, at large $c$ sector, up to $O(c^0)$ order, the flow equation becomes simple. We obtain the $O(c)$ and $O(c^0)$ order contributions to the deformed characters. The result is analogous to the $T\bar{T}$ deformation, with the one-loop contribution displaying a state-dependent rescaling of the modular parameters. The deformation effectively induces a modular parameter redefinition incorporating the deformation parameter, with modular invariance maintained for the refined modular parameters. Applying modular bootstrap trick, we derive the asymptotic density of states, which follows a Cardy-like formula. Unlike $T\bar{T}$ counterpart, the root-$T\bar{T}$ deformation exhibits no Hagedorn behaviour in our analysis.
\par
On the gravity side, the root-$T\bar{T}$ deformed CFT is dual to AdS$_3$ with certain modified boundary condition. We then reformulate the AdS$_3$ with root-$T\bar{T}$ deformed boundary condition into Chern-Simons formulation. In terms of the Chern-Simons formulation, the BTZ black hole entropy is described by the Wilson loops and the one-loop partition functions are given by Wilson spools~\cite{Castro:2023bvo,Castro:2023dxp,Bourne:2024ded}. We study the root-$T\bar{T}$ deformed AdS$_3$ and find the deformed geometry is still a quotient of hyperbolic space with a different periods. We proceed to compute the entropy of the root-$T\bar{T}$ deformed BTZ black hole and the one-loop partition function in deformed AdS$_3$, employing the Wilson loop and Wilson spool techniques, respectively. The deformed BTZ black hole entropy matches the asymptotic density of states obtained via modular bootstrap. Furthermore, we show the one-loop partition function reduces to a tracing over descendant states in the Verma module, except for the redefinition of modular parameter, which is consistent with the field theory result as well. These results provide strong evidence for the consistency of the proposed duality between root-$T\bar{T}$ deformed CFT and AdS$_3$ with root-$T\bar{T}$ deformed boundary conditions at the one-loop level.
\par
The rest of this paper is organized as follows. In section~\ref{sec:2}, we review fundamental aspects of root-$T\bar{T}$ deformation, including the deformed spectrum and its holographic description, which are closely related to this work. In section~~\ref{sec:3}, we discuss the root-$T\bar{T}$ deformed partition function, which turns out to be satisfy a flow equation. Furthermore, in the large $c$ limit, we show that the partition function becomes modular invariant under the appropriate redefinition of the modular parameter. Using modular bootstrap technique, we derive the asymptotic density of states governed by a Cardy-like formula. In section~\ref{sec:4}, we compute the one-loop partition function for AdS$_3$ with root-$T\bar{T}$ deformed boundary condition. The results precisely match the large $c$ expansion of the root-$T\bar{T}$ deformed CFT partition function. The conclusions and discussions about the future directions are given in section~\ref{sec:5}. The appendices contain some technical details about deriving the flow equation of root-$T\bar{T}$ deformed partition function.
\section{Root-$T\bar{T}$ deformation and holographic description}
\label{sec:2}
This section is a brief review of some basics about the root-$T\bar{T}$ deformation including the deformed spectrum and the holographic description, which are closely related to this work. The deformed spectrum turns out to satisfy a flow equation under some consistency assumptions. The holographic duality of the root $T\bar{T}$-deformed CFT is the AdS$_3$ with a certain modified boundary conditions, which is derived from the root-$T\bar{T}$ flow equation. Similar to the $T\bar{T}$ deformation, the root-$T\bar{T}$ deformed AdS$_3$ geometry can also be obtained through a coordinate transformation.
\subsection{Root-$T\bar{T}$ deformed spectrum}
The root-$T\bar{T}$ deformation is a generalization of the $T\bar{T}$ deformation, whose deformed action is given by the $T\bar{T}$-like flow equation
\begin{align}
\frac{\partial S_{\lambda}}{\partial \lambda}=\int d^2x\mathcal{R}_{T\bar{T}}.
\end{align}
Classically, the square-root $T\bar{T}$ operator is formally defined as 
\begin{align}
\mathcal{R}_{T\bar{T}}\equiv\sqrt{\frac{1}{2}T^{\mu\nu}T_{\mu\nu}-\frac{1}{4}T^{\mu}_{\mu}T^{\nu}_{\nu}}=\sqrt{\det\widetilde{T}},
\end{align}
where
\begin{align}
\widetilde{T}_{\mu\nu}=T_{\mu\nu}-\frac{1}{2}g_{\mu\nu}T^{\rho}_{\rho},
\end{align}
is the traceless part of the stress-energy tensor. The classical action can be obtained by solving the flow equation~\cite{Ferko:2022cix}. However, on the quantum level the square-root $T\bar{T}$ operator have not fully understood yet. For the $T\bar{T}$ deformation, the deformed spectrum can be obtained exactly because of the Zamolodchikov’s factorization formula
\begin{align}
\langle n|\mathcal{O}_{T\bar{T}}|n\rangle=\langle n|T|n\rangle\langle n|\bar{T}|n\rangle-\langle n|\Theta|n\rangle\langle n|\Theta|n\rangle.
\end{align}
In~\cite{Ebert:2023tih}, the authors assume that there exists a local operator $\mathcal{R}(x)$, which is a square root of the $T\bar{T}$ operator and satisfies the factorization 
\begin{align}
\langle n|\det(t^{\mu\nu})|n\rangle=\langle n|\mathcal{R}(x)|n\rangle\langle n|\mathcal{R}(x)|n\rangle,
\end{align}
so that the expectation value of the square-root operator reads
\begin{align}
\left\langle\mathcal{R}\right\rangle=\sqrt{\frac{1}{2}\left\langle T^{\mu\nu}\right\rangle\left\langle T_{\mu\nu}\right\rangle-\frac{1}{4}\left\langle T^{\mu}_{\mu}\right\rangle\left\langle T^{\nu}_{\nu} \right\rangle}.
\end{align}
The Zamolodchikov’s factorization of the $T\bar{T}$ operator can be straightforward recovered by taking a square. By expressing the components of the stress tensor for the theory on a cylinder of radius $R$ in terms of energies and momenta, the root-$T\bar{T}$ deformed spectrum is shown to satisfy the flow equation 
\begin{align}
\label{spectrum_flow_1}
\frac{\partial E_{n}(R,\lambda)}{\partial\lambda}=\pm\sqrt{\frac{1}{4}\left(E_{n}-R\frac{\partial E_{n}}{\partial R}\right)^{2}-P_{n}^{2}},
\end{align}
or equivalently
\begin{align}
\label{spectrum_flow_2}
\left(\frac{\partial E_{n}(R,\lambda)}{\partial\lambda}\right)^{2}-\frac{1}{4}\left(E_{n}(R,\lambda)-R\frac{\partial E_{n}(R,\lambda)}{\partial R}\right)^{2}+P_{n}^{2}=0.
\end{align}
This flow  equation can be solved if we impose certain initial condition. In this paper, we shall consider the seed theory is the 2d conformal field theory.
\paragraph{Zero-momentum case:} For the zero-momentum case $P_n=0$, the flow equation can be simplified to
\begin{align}
\frac{\partial E_n(R,\lambda)}{\partial\lambda}=\pm\frac{1}{2}\left(E_{n}(R,\lambda)-R\frac{\partial E_{n}(R,\lambda)}{\partial R}\right),
\end{align}
which lead to the following deformed spectrum
\begin{align}
\label{zero-momentum_spectrum}
E_n(R,\lambda)=E_ne^{\pm\lambda}.
\end{align}
In this case, the spectrum is just a rescaling of the original one.
\paragraph{General case:} For the general case, we consider the seed theory to be a conformal field theory whose spectrum and momentum are given by
\begin{align}
E_n=\frac{1}{R}\left(\Delta_n+\bar{\Delta}_n-\frac{c}{12}\right),\quad P_n=\frac{1}{R}\left(\Delta_n-\bar{\Delta}_n\right).
\end{align}
The flow equation can also be solved and the deformed spectrum reads
\begin{align}
\label{general_spectrum}
E_n(\lambda)=\cosh(\lambda) E_n\pm\sinh(\lambda)\sqrt{E_n^2-P_n^2} ,
\end{align}
while the momentum $P_n$ does not flow.
\par
The following comments about the deformed spectrum are in order:
\begin{itemize}
\item[•] The spectrum for different branches are both physical, because both of them can give the correct perturbation result. While for the large $\lambda$, these two branches become quite different. It should be noted that the different branches are related by changing the sign of deformation parameter. Therefore in the following, we would like to study the deformed spectrum by taking the value of deformation parameter $\lambda\in(-\infty,\infty)$ with positive sign of $\lambda$.
\item[•] The spectrum can be separated into left- and right-moving energy defined as
\begin{align}
E_n(\lambda)=E^{L}_{n}(\lambda)+E^{R}_{n}(\lambda),\quad P_n(\lambda)=E^{L}_{n}(\lambda)-E^{R}_{n}(\lambda),
\end{align}
so that the left- and right-moving energy reads
\begin{align}
\sqrt{E^{L}_{n}(\lambda)}=\sqrt{E^{L}_{n}}\cosh\left(\frac{\lambda}{2}\right)+\sqrt{E^{R}_{n}}\sinh\left(\frac{\lambda}{2}\right),\\
\sqrt{E^{R}_{n}(\lambda)}=\sqrt{E^{R}_{n}}\cosh\left(\frac{\lambda}{2}\right)+\sqrt{E^{L}_{n}}\sinh\left(\frac{\lambda}{2}\right).
\end{align}
The effect of root-$T\bar{T}$ deformation is a boost transformation between $\sqrt{E_L}$ and $\sqrt{E_R}$, which  preserves the momentum unchanged
\begin{align}
E_n^L(\lambda)-E_n^{R}(\lambda)=E_n^L-E_n^{R}.
\end{align}
\item[•] For the zero-momentum case, the deformed spectrum is just a rescaling of the original one. For the general non-zero momentum case, we expand the deformed spectrum for large $c$ CFT, which leads to 
\begin{align}
E_n(\lambda)&=E_ne^{-\lambda}+O(1/c).
\end{align}
We find, up to the $O(c^0)$ order, the deformed energy is just the undeformed energy multiply a factor $e^{-\lambda}$. While the root-$T\bar{T}$ deformed spectrum becomes complicated for the higher order.
\end{itemize}
\subsection{Root-$T\bar{T}$ deformed AdS$_3$ boundary conditions}
In~\citep{Ebert:2023tih}, it was proposed the root-$T\bar{T}$ deformed CFT corresponds to AdS$_3$ with modified boundary condition. The deformed bulk geometry can be obtained through a coordinate transformation. We start from the general AdS$_3$ solution, which in the Fefferman-Graham gauge can be written as
\begin{align}
ds^2 = \frac{dr^2}{r^2} + \left( r^2 g^{(0)}_{ij}(x) + g^{(2)}_{ij}(x) + \frac{g^{(4)}_{ij}(x)}{r^2} \right) dx^{i} dx^{j}.
\end{align}
In the standard AdS/CFT correspondence, $g^{(0)}_{ij}$ is the metric where the boundary field theory living in and $g^{(2)}_{ij}$ is related to the expectation value of the stress tensor
\begin{align}
g_{ij}^{(2)} = 2\left( T_{ij} - g_{ij}^{(0)} T_{k}^{k} \right).
\end{align}
By solving the root-$T\bar{T}$ flow equation with consistency conditions~\cite{Ebert:2023tih}, one can obtain the root-$T\bar{T}$ deformed metric and stress tensor. In terms of the AdS$_3$ quantities, the proper root-$T\bar{T}$ deformed boundary metric and stress tensor are given by
\begin{align}
\gamma_{ij}^{(\lambda)}&=\cosh(\lambda)g_{ij}^{(0)}+\frac{\sinh(\lambda)}{\mathcal{R}^{(0)}}\widetilde{T}_{ij}^{(0)},\\
\widetilde{T}_{ij}^{(\lambda)}&=\cosh(\lambda)\widetilde{T}_{ij}^{(0)}+\sinh(\lambda)\mathcal{R}^{(0)}g_{ij}^{(0)},
\end{align}
where $\widetilde{T}_{ij}^{(0)}$ is the traceless part of the stress tensor and $\mathcal{R}^{(0)}$ is the square root of $\widetilde{T}_{ij}^{(0)}$ determinant.  
\par
We consider the Ba\~nados geometry~\cite{Banados:1998gg}, namely the AdS$_3$ solution with Brown-Henneaux boundary condition~\cite{Brown:1986nw}, which in Fefferman-Graham gauge becomes
\begin{align}
\label{banados}
ds^2=\frac{dr^2}{r^2}+r^2dudv+\mathcal{L}(u)du^2+\overline{\mathcal{L}}(v)dv^2+\frac{1}{r^2}\mathcal{L}(u)\overline{\mathcal{L}}(v)dudv,
\end{align}
where $\mathcal{L}(u)$ and $\bar{\mathcal{L}}(v)$ are holomorphic and anti-holomorphic function, respectively. In this case, the stress tensor is traceless $\widetilde{T}_{ij}=T_{ij}$ and the square root $T\bar{T}$-operator equals
\begin{align}
\mathcal{R}^{(0)}=\sqrt{\mathcal{L}(u)\overline{\mathcal{L}}(v)}.
\end{align}
The deformed Ba\~nados geometry can be obtained through a dynamical change of coordinate. We first replace the parameter $\mathcal{L},\bar{\mathcal{L}}$ by $\mathcal{L}_{\lambda},\bar{\mathcal{L}}_{\lambda}$, then do the following coordinate transformation
\begin{align}
\label{root_ct_1}
du&=\left(\cosh\frac{\lambda}{2}\right)dU-\sqrt{\frac{\overline{\mathcal{L}}_{\lambda}}{\mathcal{L}_{\lambda}}}\left(\sinh\frac{\lambda}{2}\right)dV,\\
\label{root_ct_2}
dv&=\left(\cosh\frac{\lambda}{2}\right)dV-\sqrt{\frac{\mathcal{L}_{\lambda}}{\overline{\mathcal{L}}_{\lambda}}}\left(\sinh\frac{\lambda}{2}\right)dU,
\end{align}
where $\mathcal{L}_{\lambda},\bar{\mathcal{L}}_{\lambda}$ can be fixed by this deformation is smooth and preserves degeneracy of states of the boundary theory which means the black hole horizon area is unchanged, as well as root-$T\bar{T}$ deformation does not affect the momentum in the boundary field theory
\begin{align}
\mathcal{L}_{\lambda}&=\left(\sqrt{\mathcal{L}(u)}\cosh\frac{\lambda}{2}+\sqrt{\overline{\mathcal{L}}(v)}\sinh\frac{\lambda}{2}\right)^{2},\\
\overline{\mathcal{L}}_{\lambda}&=\left(\sqrt{\overline{\mathcal{L}}(v)}\cosh\frac{\lambda}{2}+\sqrt{\mathcal{L}(u)}\sinh\frac{\lambda}{2}\right)^{2}.
\end{align}
This is similar to the result for $T\bar{T}$ deformation. This holographic proposal have been verified on the classical action and the deformed spectrum. In the remaining of this work, we would like to study the partition function of root-$T\bar{T}$ deformed CFT from both boundary field theory and holography.
\section{Root-$T\bar{T}$ deformed partition functions}
\label{sec:3}
For the conformal field theory, the torus partition function can be obtained by tracing over the Verma module of Virasoro algebra
\begin{align}
Z_0(\tau,\bar{\tau})=\mathrm{Tr}\left(q^{L_0-\frac{c}{24}}\bar{q}^{\bar{L}_0-\frac{c}{24}}\right)=\sum_{n=0}^{\infty}e^{-2\pi R \tau_2E_n+2\pi i\tau_1R P_n}.
\end{align}
The CFT partition function is modular invariant
\begin{align}
Z_0\left(\frac{a\tau+b}{c\tau+d},\frac{a\bar{\tau}+b}{c\bar{\tau}+d}\right)=Z_0(\tau,\bar{\tau}),\quad ad-bc=1,\quad a,b,c,d\in \mathbb{Z},
\end{align}
where the modular parameter is given by
\begin{align}
\tau=\tau_1+i\tau_2,\quad \bar{\tau}=\tau_1-i\tau_2.
\end{align}
Similarly, the root-$T\bar{T}$ deformed partition function on a torus can also be calculated by replacing the spectrum with the deformed one, which can be written as 
\begin{align}
Z(\tau,\bar{\tau}|\lambda)=\sum_{n=0}^{\infty}e^{-2 \pi  R \tau_2 E_n(R,\lambda )+2 \pi  i R \tau_1 P_n}.
\end{align}
In what following, we shall study root-$T\bar{T}$ deformations of conformal field theories on the torus. We first show the deformed partition function satisfy a flow equation and is modular invariant after redefining the modular parameter for the zero-momentum case. We then generalize the result to the more general case especially for the large central charge sector. 
\subsection{Zero-momentum case}
For the zero-momentum case, the deformed spectrum is a rescaling of the original one~\eqref{zero-momentum_spectrum}. The deformed partition function can be written as
\begin{align}
\mathcal{Z}(\tau,\bar{\tau}|\lambda)=\sum_{n=0}^{\infty}e^{-2 \pi  R \tau_2 E_ne^{\lambda}}.
\end{align}
We then do the perturbation expansion of the partition function with respect to the deformation parameter
\begin{align}
\mathcal{Z}(\tau,\bar{\tau}|\lambda)=\sum_{n=0}^\infty\lambda^{n}Z_n(\tau,\bar{\tau}) .
\end{align}
The first few orders read
\begin{align}
Z_0&=\sum_{n=0}^{\infty}e^{-2 \pi  R \tau_2 E_n},\\
Z_1&=\sum_{n=0}^{\infty}-2 \pi  E_n R \tau _2 e^{-2 \pi  E_n R \tau _2},\\
Z_2&=\sum_{n=0}^{\infty}\pi  E_n R \tau _2 \left(2 \pi  E_n R \tau _2-1\right)e^{-2 \pi  E_n R \tau _2} ,\\
&\cdots\nonumber
\end{align}
There is a polynomial of the energy for each order in perturbation theory. The polynomials are related to derivatives of $Z_0$ with respect to the modular parameter. We find the following replacement rule
\begin{align}
2\pi RE_n\mapsto-\partial_{\tau_2}.
\end{align}
Then the first few order can be written as 
\begin{align}
Z_1=\tau_2\partial_{\tau_2}Z_0,\quad Z_2=\frac{1}{2}\tau_2\partial_{\tau_2}Z_1,\quad \quad Z_3=\frac{1}{3}\tau_2\partial_{\tau_2}Z_2,\quad \cdots
\end{align}
and generally we have
\begin{align}
\label{RR-1}
Z_p=\frac{1}{p}\tau_2\partial_{\tau_2}Z_{p-1}.
\end{align}
Therefore, the deformed partition functions satisfy the flow equation
\begin{align}
\label{flow-1}
\partial_{\lambda}\mathcal{Z}(\tau,\bar{\tau}|\lambda)=\tau_2\partial_{\tau_2}\mathcal{Z}(\tau,\bar{\tau}|\lambda),
\end{align}
with the initial condition
\begin{align}
\mathcal{Z}(\tau,\bar{\tau}|0)=Z_0(\tau,\tau).
\end{align}
We have to point out that the deformed partition function is \emph{not modular invariant}. In fact, according to the recurrence relation~\eqref{RR-1}, we start from a modular invariant partition function $Z_0$, while the other order corrections of the partition functions are \emph{not modular invariant}. Since the spectrum is a rescaling of the original one, the partition function can be written as
\begin{align}
\mathcal{Z}(\tau,\bar{\tau}|\lambda)=Z_0(\tau_{\lambda},\bar{\tau}_{\lambda}),\quad \text{with}\quad 
\tau_{\lambda}=ie^{\lambda}\tau_2,\quad \bar{\tau}_{\lambda}=ie^{\lambda}\tau_2.
\end{align}
The deformed partition function corresponds to the redefinition of the modular parameters, maintaining modular invariance under the refined modular parameters.
\subsection{General case}
\label{sec:3.2}
From the zero-momentum case, we learn that the partition function is not the usual modular invariant. The operator $\tau_2\partial_{\tau_2}$ is not Ramanujan-Serre derivative, which maps the modular form to another modular form, see appendix~\ref{app:1} for details. We infer that the general partition function is also not modular invariant. We can not follow the method in~\cite{Datta:2018thy, Aharony:2018bad} using the modular properties to find the flow equation for the general case. While for the large $c$ sector, up to $c^{0}$ order contribution, the spectrum is a rescaling of the original one. In this case, the contribution from the descendant levels is linear just like the CFT spectrum. In this subsection, we begin by examining the large central charge sector, then extend our analysis to general case.
\paragraph{Large central charge sector}
In the large $c$ limit, the spectrum is given by 
\begin{align}
E_n(R,\lambda)&=E_ne^{-\lambda}+O(1/c),
\end{align}
which is just a rescaling of the original one. Therefore we learn that at large $c$ sector, up to the $O(c^0)$ order contribution, the deformed theory is described by a conformal field theory. The partition function reads
\begin{align}
\mathcal{Z}(\tau,\bar{\tau}|\lambda)=\sum_{n=0}^{\infty}e^{-2 \pi  R \tau_2  E_n(R)e^{-\lambda}+2\pi i R \tau_1 P_n}=Z_0(\tau_{\lambda},\bar{\tau}_{\lambda}),
\end{align}
where the deformed modular parameters are
\begin{align}
\tau_{\lambda}=\tau_1+i\tau_2e^{-\lambda},\quad \bar{\tau}_{\lambda}=\tau_1-i\tau_2e^{-\lambda},
\end{align}
or equivalent we have
\begin{align}
\label{new-modparameter}
\tau_{\lambda}=\frac{1}{2}\Big(\tau+\bar{\tau}+e^{-\lambda}(\tau-\bar{\tau})\Big),\quad \bar{\tau}_{\lambda}=\frac{1}{2}\Big(\tau+\bar{\tau}+e^{-\lambda}(\bar{\tau}-\tau)\Big).
\end{align}
The deformed partition function is a replacement of the modular parameter, which is \emph{not modular invariant} generally. However, it is obviously that the deformed partition function has the following modular properties 
\begin{align}
Z_0\left(\frac{a\tau_{\lambda}+b}{c\tau_{\lambda}+d},\frac{a\bar{\tau}_{\lambda}+b}{c\bar{\tau}_{\lambda}+d}\right)=Z_0(\tau_{\lambda},\bar{\tau}_{\lambda}),\quad ad-bc=1\quad  a,b,c,d\in \mathbb{Z}, 
\end{align}  
which means the deformed partition function at large $c$ sector is invariant under the modular transformation of deformed modular parameter. Following the similar procedure in zero-momentum case, one can find the partition functions satisfy the flow equation
\begin{align}
\partial_{\lambda}\mathcal{Z}(\tau,\bar{\tau}|\lambda)=-\tau_2\partial_{\tau_2}\mathcal{Z}(\tau,\bar{\tau}|\lambda).
\end{align}
\par
The modular invariant CFT partition function can be written as summing of characters
\begin{align}
Z(\tau,\bar{\tau})=\sum_{h,\bar{h}}\chi_{h}(\tau)\chi_{\bar{h}}(\bar{\tau}) .
\end{align}
The characters describe the degeneracy for a given level $(n_L,n_R)$, which is given by the product of integer partitions $p(n_L)$ and $p(n_R)$. We can therefore write the product of left and right-moving characters as follows
\begin{align}
\Xi(\tau,\bar{\tau})=&|\chi(\tau)|^2=e^{2\pi\tau_2\frac{c}{12}}\sum_{n_L,n_R=0}^\infty p(n_L)p(n_R)e^{-2\pi\tau_2(n_L+n_R)}e^{2\pi i\tau_1(n_L-n_R)}\nonumber\\
=&e^{2\pi\tau_{2}\frac{c}{12}}\left|\prod_{n=1}^{\infty}\frac{1}{1-e^{2\pi i\tau n}}\right|^{2}.
\end{align}
Fortunately, the deformed spectrum is a rescaling of the CFT spectrum and the degeneracy remains unchanged. We evaluate the deformed character follow the same technique used in CFT
\begin{align}
\label{rootttbar_character}
\Xi(\tau,\bar{\tau}|\lambda)=&e^{2\pi\tau_2\frac{ce^{-\lambda}}{12}}\sum_{n_L,n_R=0}^\infty p(n_L)p(n_R)e^{-2\pi\tau_2(n_L+n_R)e^{-\lambda}}e^{2\pi i\tau_1(n_L-n_R)}\nonumber\\
=&e^{2\pi\tau_{2}\frac{ce^{-\lambda}}{12}}\left|\prod_{n=1}^{\infty}\frac{1}{1-e^{2\pi i\tau_1 n-2\pi\tau_2 ne^{-\lambda}}}\right|^{2}.
\end{align}
For the vacuum character, the above product starts from $n=2$, if we take into account the null states. After redefining the modular parameter~\eqref{new-modparameter}, the characters can be written as 
\begin{align}
\Xi(\tau,\lambda)=|q_{\lambda}|^{-2k}\prod_{n=2}^{\infty}\frac{1}{|1-q_{\lambda}^n|^2},
\end{align}
where
\begin{align}
q_{\lambda}=e^{2\pi i\tau_{\lambda}},\quad \bar{q}_{\lambda}=e^{-2\pi i\bar{\tau}_{\lambda}} .
\end{align}
Similar properties also find in $T\bar{T}$ deformation~\cite{Datta:2021kha}, which is described by a $T\bar{T}$ deformed modular form~\cite{Cardy:2022mhn}. While for the higher order contribution, the deformed partition function becomes more complicated.
\paragraph{Beyond large central charge}
For the higher orders, the deformed left- and right-moving spectrum are coupled which may not be described by a conformal field theory. The partition function still have the following expansion
\begin{align}
\mathcal{Z}(\tau,\bar{\tau}|\lambda)=\sum_{n=0}^\infty\lambda^{n}Z_n(\tau,\bar{\tau}) .
\end{align}
The first few orders read
\begin{align}
Z_0&=\sum_{n=0}^{\infty}e^{2\pi i R \tau_1P_n-2\pi R\tau_2E_n},\\
Z_1&=\sum_{n=0}^{\infty}\Big(-2 \pi R \tau_2 \Big)\sqrt{E_n^2-P_n^2}e^{2\pi i R \tau_1P_n-2\pi R\tau_2E_n},\\
Z_2&=\sum_{n=0}^{\infty}\Big(2 \pi ^2 R^2 \left(E_n^2-P_n^2\right) \tau_2^2-\pi R E_n\tau_2\Big)e^{2\pi i R \tau_1P_n-2\pi R\tau_2E_n},\\
Z_3&=\sum_{n=0}^{\infty}\left(-\frac{4}{3}\pi ^3  R^3(E_n^2-P_n^2) \tau _2^3+2\pi^2 R^2 E_n \tau_2^2-\frac{1}{3}\pi R\tau_2\right)\sqrt{E_n^2-P_n^2} e^{2\pi i R \tau_1P_n-2\pi R\tau_2E_n},\\
&\cdots\nonumber
\end{align}
We find the odd orders are related to the square root term, while the even orders are a polynomial of energy and momentum. The perturbation partition functions can be converted into actions of derivatives on the CFT partition function, except for the square root term. We have the following replacements rule
\begin{align}
\label{rep_1}
2\pi RE_n\mapsto-\partial_{\tau_2}=-i(\partial_\tau-\partial_{\bar{\tau}}),\quad2\pi iRP_n\mapsto\partial_{\tau_1}=\partial_\tau+\partial_{\bar{\tau}},
\end{align}
as well as 
\begin{align}
\label{rep_2}
(2\pi R)^2\left(E_n^2-P_n^2\right)\mapsto 4\partial_{\tau}\partial_{\bar{\tau}}.
\end{align}
We therefore rewrite the first few order expansion of the partition function as 
\begin{align}
Z_1&=-\tau_2\widehat{\mathcal{R}}Z_0,\\
Z_2&=\left[2\tau_2^2\partial_{\tau}\partial_{\bar{\tau}}+\frac{1}{2}\tau_2\partial_{\tau_2}\right]Z_0,\\
Z_3&=\left[-\frac{2}{3}\tau_2^3\partial_{\tau}\partial_{\bar{\tau}}-\frac{1}{2}\tau_2^2\partial_{\tau_2}-\frac{1}{6}\tau_2\right]\widehat{\mathcal{R}}Z_0,\\
&\cdots\nonumber
\end{align}
where
\begin{align}
\widehat{\mathcal{R}}Z_0=\sum_{n=0}^{\infty}2 \pi R \sqrt{E_n^2-P_n^2}e^{2\pi i R \tau_1P_n-2\pi R\tau_2E_n}.
\end{align}
More generally, we conclude
\begin{align}
Z_p=
\left\{
  \renewcommand{\arraystretch}{2}  
     \begin{array}{lr}  
\left(\sum_{m=1}^{p}\tau_2^m\widehat{\mathcal{O}}_m^{(p)}\right)Z_0(\tau,\bar{\tau}),&\text{for even $p$}\\
\left(\sum_{m=1}^{p}\tau_2^m\widehat{\mathcal{O}}_m^{(p)}\right)\widehat{\mathcal{R}}Z_0(\tau,\bar{\tau}),&\text{for odd $p$}
\end{array}
\right.
\end{align}
which means the perturbation terms can be written as a polynomial of $\tau_2$ with a differential operator $\widehat{\mathcal{O}}_n^{(p)}(\partial_\tau,\partial_{\bar{\tau}})$ as coefficients act on the original partition function. The information about the expansion of spectrum are encode in the operators.
\par
For the $T\bar{T}$ deformation, the recurrence relation for the perturbation was obtained using the modular properties of the partition function. However, the odd orders are related to a square root term, which cannot been written in terms of a regular differential operator in the root-$T\bar{T}$ deformation. In fact, we can also define the square root operator see appendix~\ref{app:1} for details. Another important observation about the expansion of the partition function is that the square root term only appear in odd order perturbations, and the even order terms are well-defined. In order to avoid the square root term, we consider the second order recurrence relation. Therefore, we have to consider the recursion relation between $Z_{p+2}$ and $Z_p$. The $p$-order correction of the partition function is a $p$-order polynomial of $\tau_2$ with the coefficient to be a function of $E_n$ and $E^2_n-P_n^2$, which can be transformed to a differential operator using the replacement rule~\eqref{rep_1} and \eqref{rep_2}. There is no constant term as well. According to above observation and analysis, we make the following ansatz for the recurrence relation
\begin{align}
Z_{p+2}=\mathcal{D}_{p}Z_p,
\end{align}
where
\begin{align}
\label{D-ansatz}
\mathcal{D}_{p}=&A_{p}\tau_2^2\partial_{\tau}\partial_{\bar{\tau}}+B_{p}\tau_2\partial_{\tau_2}.
\end{align}
Note that the first part of this operator is Ramanujan-Serre derivative, which maps one modular form to another modular form, while the second operator is not a Ramanujan-Serre derivative, see~\eqref{app:1} for details. Therefore the differential operator do not preserve the modular form. Moreover, we find 
\begin{align}
\mathcal{D}_{p}\tau_2^n=A_{p}\tau_2^{n+2}\partial_{\tau}\partial_{\bar{\tau}}+B_{p}\tau_2^{n+1}\partial_{\tau_2}+\left(\frac{n(n-1)}{4}A_{p}+nB_{p}\right)\tau_2^n.
\end{align}
The coefficient $A_p,B_p$ in this ansatz~\eqref{D-ansatz} can be fixed by considering the highest and lowest order of $\tau_2$. The highest order of $\tau_2$ is given by
\begin{align}
\mathcal{O}_{p}^{(p)}=
\left\{
\renewcommand{\arraystretch}{2}
\begin{array}{lr}
\frac{2^{p}}{p!}\left(\partial_{\tau}\partial_{\bar{\tau}}\right)^{p/2},& \text{for even $p$}\\
-\frac{2^{p}}{p!}\left(\partial_{\tau}\partial_{\bar{\tau}}\right)^{(p-1)/2},& \text{for odd $p$}
\end{array}
\right.
\end{align}
Comparing the operator coefficient, we get
\begin{align}
A_p=\frac{4}{(p+1)(p+2)},
\end{align}
while the lowest $\tau_2$ order is given by
\begin{align}
\mathcal{O}_{1}^{(p)}=
\left\{
\renewcommand{\arraystretch}{2}
\begin{array}{lr}
\frac{1}{p!}\partial_{\tau_2},& \text{for even $p$}\\
-\frac{1}{p!},& \text{for odd $p$}
\end{array}
\right.
\end{align}
So that we have
\begin{align}
B_{p}=\frac{1}{(p+1)(p+2)}.
\end{align}
Therefore, the recurrence relation becomes
\begin{align}
\label{partition_flow_equation}
Z_{p+2}=\frac{1}{(p+1)(p+2)}\Big(4\tau_2^2\left(\partial_{\tau}\partial_{\bar{\tau}}\right)+\tau_2\partial_{\tau_2}\Big)Z_p,
\end{align} 
which holds for both even and odd $p$. The only difference between the odd terms and even terms is the initial condition. The recurrence relation implies the flow equation for the deformed partition function
\begin{align}
\partial_{\lambda}^{2}\mathcal{Z}(\tau,\bar{\tau}|\lambda)=\left(4\tau_{2}^{2}\left(\partial_{\tau}\partial_{{\bar{\tau}}}\right)+\tau_{2}\partial_{{\tau_{2}}}\right)\mathcal{Z}(\tau,\bar{\tau}|\lambda),
\end{align}
with the initial condition
\begin{align}
\label{initial_condition}
\mathcal{Z}(\tau,\bar{\tau}|\lambda)\Big|_{\lambda=0}=Z_0,\quad \partial_{\lambda}\mathcal{Z}(\tau,\bar{\tau}|\lambda)\Big|_{\lambda=0}=Z_1.
\end{align}
This is a second order differential equation, which is different from the $T\bar{T}$ deformation. In our case, the information of root-$T\bar{T}$ deformation is also encoded in the initial condition~\eqref{initial_condition}. This flow equation for the partition function is also obtained in~\cite{Ferko:2023iha}. It looks like we have already taken a square of the differential equation similar to the flow equation of the deformed spectrum~\eqref{spectrum_flow_1} and ~\eqref{spectrum_flow_2}. However, we do not find the modular properties of the root-$T\bar{T}$ deformed partition function from the flow equation. Notably, we may alternatively perform a Laurent expansion in terms of the parameter $t=e^{-\lambda}$, see Appendix~\ref{app:2} for details.
\subsection{Asymptotic density of state}
We consider the zero-momentum case, and the partition function has the following modular properties
\begin{align}
\mathcal{Z}\left(\frac{a\tau_{\lambda}+b}{c\tau_{\lambda}+d},\frac{a\bar{\tau}_{\lambda}+b}{c\bar{\tau}_{\lambda}+d}\right)=\mathcal{Z}(\tau_{\lambda},\bar{\tau}_{\lambda}).
\end{align} 
For a rectangular torus with a purely imaginary modular parameter, we choose the modular parameter to be $\tau=i\beta/R$, which implies that
\begin{align}
\tau_{\lambda}=\frac{i\beta}{2\pi R}e^{\lambda},\quad \bar{\tau}_{\lambda}=-\frac{i\beta}{2\pi R}e^{\lambda}.
\end{align}
In this case, the partition function simplifies to
\begin{align}
\mathcal{Z}(\beta,\lambda)=\sum_{n=0}^{\infty}e^{-2\pi\beta E_{n}(R,\lambda)}.
\end{align}
For the low temperature limit, the partition function is dominated by the ground state
\begin{align}
E_0(R,\lambda)=-\frac{c}{12R}e^{\lambda},
\end{align}
The partition function becomes
\begin{align}
\lim_{\beta\to\infty}\mathcal{Z}(\beta,\lambda)=\exp\left(\frac{\pi\beta c}{6R}e^{\lambda}\right).
\end{align}
Now we perform the $\mathcal{S}$-modular transformation of the deformed modular parameters, which is equivalent to 
\begin{align}
\frac{\beta}{R}e^{\lambda}\to \frac{R}{\beta}e^{-\lambda}.
\end{align}
Physically, the $\mathcal{S}$-modular transformation maps the low temperature limit to the high temperature limit. In the high energy limit, almost all the state contributes. We can define the following spectral density
\begin{align}
\rho(E)=\sum_{n}\delta(E-E_n(R,\lambda)).
\end{align}
Then the partition function at high temperature limit can be written as 
\begin{align}
\mathcal{Z}(\beta,\lambda)=\int_{E_0(R,\lambda)}^{\infty}dE \rho(E)e^{-2\pi \beta E}=\exp\left(\frac{\pi R c}{6\beta}e^{-\lambda}\right),
\end{align}
so that
\begin{align}
\rho(E)=\int d\beta \exp\left(\beta E-\frac{\pi R c}{6\beta}e^{-\lambda}\right).
\end{align}
Under saddle point approximation, we obtain the asymptotic density of state
\begin{align}
\label{cardy_like}
\rho(E)=\exp\left(\sqrt{\frac{2 \pi c R E }{3}} e^{-\frac{\lambda}{2}}\right),
\end{align}
which is the Cardy formula for 2d CFT except for the factor $e^{-\frac{\lambda}{2}}$. This result agrees with the holographic prediction in~\cite{Ebert:2023tih}. For the large $c$ sector, the partition function have the similar properties since the spectrum is a rescaling of the original one. However, for the non-zero momentum case, we do not find the modular properties of the partition function generally.  
\section{Holographic calculation}
\label{sec:4}
In the context of AdS/CFT, the conformal field theory at large $c$ limit dual to the semi-classical limit of the AdS$_3$ gravity. The $O(c^0)$ order contribution on the quantum field theory is dual to the one-loop correction of quantum gravity. In this section, we shall check the root-$T\bar{T}$ deformed holographic duality proposal by calculating the one-loop partition function of root-$T\bar{T}$ deformed AdS$_3$. 
\par
We mainly focus on the BTZ black hole, in this case $\mathcal{L}_0$ and $\bar{\mathcal{L}}_0$ in Ba\~nados geometry~\eqref{banados} become the constants which are related to the mass and angular momentum of BTZ black hole 
\begin{align}
\mathcal{L}_0&=\frac{M+J}{2},\quad \overline{\mathcal{L}}_0=\frac{M-J}{2}.
\end{align}
The AdS$_3$ should be locally the hyperbolic space $\mathbb{H}_3$ or a quotient of $\mathbb{H}_3$ by some discrete group $\Gamma$~\cite{Kraus:2006wn}. In fact, one can perform the following coordinate transformation 
\begin{align}
\label{ct_1}
y=&\frac{2\sqrt[4]{\mathcal{L}_0\overline{\mathcal{L}}_0}r}{r^2+\sqrt{\mathcal{L}_0\overline{\mathcal{L}}_0}}e^{\sqrt{\mathcal{L}_0}u+\sqrt{\overline{\mathcal{L}}_0}v},\\
\label{ct_2}
\xi=&\frac{r^2-\sqrt{\mathcal{L}_0\overline{\mathcal{L}}_0}}{r^2+\sqrt{\mathcal{L}_0\overline{\mathcal{L}}_0}}e^{2 \sqrt{\mathcal{L}_0}u},\\
\label{ct_3}
\bar{\xi}=&\frac{r^2-\sqrt{\mathcal{L}_0\overline{\mathcal{L}}_0}}{r^2+\sqrt{\mathcal{L}_0\overline{\mathcal{L}}_0}}e^{2 \sqrt{\overline{\mathcal{L}}_0}v},
\end{align}
where $u=\theta'+it',v=\theta'-it'$. So that the metric becomes the Poincar\'e patch
\begin{align}
ds^2=\frac{dy^2+d\xi d\bar{\xi}}{y^2},
\end{align}
where $y>0$ and $\xi$ is a complex coordinate. We have to point out that the coordinate transformation just holds for the region outside the horizon of BTZ black hole. The Poincar\'e patch can be written into the line element on $\text{SL}(2,\mathbb{C})$
\begin{align}
\label{g-metric}
ds^2=\frac{1}{2}\mathrm{Tr}(g^{-1}dgg^{-1}dg),\quad g=\left(\begin{array}{cc}y+{\xi\bar{\xi}}/{y}&{\xi}/{y}\\{\bar{\xi}}/{y}&{1}/{y}\end{array}\right)\in \text{SL}(2,\mathbb{C}),
\end{align}
which has the following discrete group of isometry generated by $\gamma$
\begin{align}
g\to \gamma g\gamma^{\dagger},\quad \gamma=\left(\begin{array}{cc}e^{i\pi\tau_0}&0\\ 0 &e^{-i\pi\tau_0}\end{array}\right)\in \text{SL}(2,\mathbb{C}).
\end{align}
From the exponential part of the coordinate transformations \eqref{ct_1}-\eqref{ct_3}, one can find the identifications of the coordinates 
\begin{align}
u\sim u+\frac{n\pi i}{\sqrt{\mathcal{L}_0}},\quad v\sim v+\frac{n\pi i}{\sqrt{\overline{\mathcal{L}}_0}}.
\end{align}
The discrete group of isometry implies the modular parameters are
\begin{align}
-\frac{1}{\tau_0}=\frac{i}{2\sqrt{\mathcal{L}_0}},\quad -\frac{1}{\bar{\tau}_0}=-\frac{i}{2\sqrt{\overline{\mathcal{L}}_0}}.
\end{align}
The identifications of $\theta'$ and $t'$ coordinate becomes
\begin{align}
\theta'\sim\theta'+2\pi,\quad \theta'\sim \theta'+\beta'_1,\quad t'\sim t'+\beta'_2.
\end{align}
The modular parameter is defined as the ratio of these two periods
\begin{align}
-\frac{1}{\tau_0}=\frac{\beta_1+i\beta_2}{2\pi},\quad -\frac{1}{\bar{\tau}_0}=\frac{\beta'_1-i\beta'_2}{2\pi}.
\end{align}
This is the quotient space structure of BTZ black hole. The similar structure was found in $T\bar{T}$-deformed BTZ black hole~\cite{He:2024pbp}. We then will show the root-$T\bar{T}$ deformed BTZ black hole is also a quotient space structure with different modular parameters.
\par
We have learned that the root-$T\bar{T}$ deformed BTZ black hole solution can be obtained by replacing the parameter $\mathcal{L}_0,\overline{\mathcal{L}}_0$ with $\mathcal{L}_{\lambda},\overline{\mathcal{L}}_{\lambda}$ then doing the coordinate transformation~\eqref{root_ct_1} and~\eqref{root_ct_2}. Combining with the coordinate transformation~\eqref{ct_1}-\eqref{ct_3}, we can also transform the root-$T\bar{T}$ deformed BTZ black hole metric into the Poincar\'e patch
\begin{align}
y=&\frac{2\sqrt[4]{\mathcal{L}_{\lambda}\overline{\mathcal{L}}_{\lambda}}r}{r^2+\sqrt{\mathcal{L}_{\lambda}\overline{\mathcal{L}}_{\lambda}}}\exp\left[e^{-\frac{\lambda}{2}}\left(\sqrt{\mathcal{L}_{\lambda}}U+\sqrt{\overline{\mathcal{L}}_{\lambda}}V\right)\right],\\
\xi=&\frac{r^2-\sqrt{\mathcal{L}_{\lambda}\overline{\mathcal{L}}_{\lambda}}}{r^2+\sqrt{\mathcal{L}_{\lambda}\overline{\mathcal{L}}_{\lambda}}}\exp\left[2 \left(\sqrt{\mathcal{L}_{\lambda}}\cosh\left(\frac{\lambda }{2}\right)U-\sqrt{\overline{\mathcal{L}}_{\lambda}}\sinh \left(\frac{\lambda }{2}\right)V\right)\right],\\
\bar{\xi}=&\frac{r^2-\sqrt{\mathcal{L}_{\lambda}\overline{\mathcal{L}}_{\lambda}}}{r^2+\sqrt{\mathcal{L}_{\lambda}\overline{\mathcal{L}}_{\lambda}}}\exp\left[2 \left(\sqrt{\overline{\mathcal{L}}_{\lambda}}\cosh \left(\frac{\lambda }{2}\right)V-\sqrt{\mathcal{L}_{\lambda}}\sinh \left(\frac{\lambda }{2}\right)U\right)\right].
\end{align}
According to the periodicity of $(u,v)$ in BTZ black hole and the root-$T\bar{T}$ deformation coordinate transformation~\eqref{root_ct_1}-\eqref{root_ct_2}, we find the periodicity in the deformed BTZ black hole
\begin{align}
U\sim U+\frac{in\pi e^{-\frac{\lambda}{2}}}{\sqrt{\mathcal{L}_{\lambda}}},\quad V\sim V+\frac{in\pi e^{-\frac{\lambda}{2}}}{\sqrt{\mathcal{L}_{\lambda}}},
\end{align}
or
\begin{align}
\theta\sim \theta+2\pi,\quad \theta\sim \theta+\beta_1,\quad t\sim t+\beta_2,
\end{align}
where
\begin{align}
\beta_1=\frac{i \pi  e^{-\frac{\lambda}{2}}}{2}\left(\frac{1}{\sqrt{\mathcal{L}_{\lambda}}}-\frac{1}{\sqrt{\overline{\mathcal{L}}_{\lambda}}}\right),\quad \beta_2=-\frac{\pi e^{-\frac{\lambda}{2}}}{2}\left(\frac{1}{\sqrt{\mathcal{L}_{\lambda}}}+\frac{1}{\sqrt{\overline{\mathcal{L}}_{\lambda}}}\right).
\end{align}
Therefore the root-$T\bar{T}$ deformed BTZ black hole is a quotient of hyperbolic space. The modular parameters are 
\begin{align}
-\frac{1}{\tau}=\frac{ie^{-\frac{\lambda}{2}}}{2\sqrt{\mathcal{L}_{\lambda}}},\quad -\frac{1}{\bar{\tau}}=-\frac{ie^{-\frac{\lambda}{2}}}{2\sqrt{\overline{\mathcal{L}}_{\lambda}}}.
\end{align}
\subsection{Chern-Simons formulation of root-$T\bar{T}$ deformed AdS$_3$}
The AdS$_3$ gravity has no local degrees of freedom, which is purely topological and can be formulated as a Chern-Simons theory~\cite{Witten:1988hc}. The associated isometry group is $\mathrm{SO}(2,2)\simeq \mathrm{SL}(2,\mathbb{R})\times \mathrm{SL}(2,\mathbb{R})$. The Chern-Simons formulation is powerful to calculate the quantities in AdS$_3$ using the gauge theory technique. For example, the Wilson loop corresponds to the BTZ black hole entropy, Wilson lines are related to the holographic entanglement entropy~\cite{Ammon:2013hba,He:2023xnb}. The one-loop partition function can also calculated by using the Wilson spool proposal. In this subsection, we would like to transform the root-$T\bar{T}$ deformed AdS$_3$ in Chern-Simons formulation.
\par 
Then Einstein-Hilbert action can be written as the difference of two copies $\mathrm{SL}(2,\mathbb{R})$ Chern-Simons theories
\begin{align}
I_{\mathrm{EH}}[e, \omega]& = I_{CS}[A]- I_{CS}[\bar{A}],
\end{align}
where the Chern-Simons action is 
\begin{align}
I_{C S}[A]&=\frac{k}{4 \pi} \int_{\mathcal{M}} \text{Tr}\left(A \wedge d A+\frac{2}{3} A \wedge A \wedge A\right),\quad k=\frac{1}{4G}.
\end{align}
The gauge fields $A$ and $\bar{A}$ are valued in $\mathfrak{sl}(2,\mathbb{R})$, which are related to the gravitational vielbein and spin connection
\begin{align}
A=\left(\omega^a+e^a\right)L_a,\quad \bar{A}=\left(\omega^a-e^a\right)L_a.
\end{align}
The $L_a$ are $\mathfrak{sl}(2,\mathbb{R})$ generators, they satisfy the commutation relations
\begin{align}
[L_a,L_b]=(a-b)L_{a+b},\quad a,b\in\{0,\pm1\}.
\end{align} 
The non-zero components of non-degenerate bilinear form are given by
\begin{align}
\label{bilinear}
\operatorname{Tr}(L_0L_0)=\dfrac{1}{2},\quad\operatorname{Tr}(L_{-1}L_1)=\operatorname{Tr}(L_1L_{-1})=-1.
\end{align}
Variation of the action leads to the equations of motion  
\begin{align} 
F\equiv dA+A\wedge A=0,\quad \bar{F}\equiv d\bar{A}+\bar{A}\wedge\bar{A}=0,
\end{align}
which are equivalent to the gravitational field equation and torsion free equation. The AdS$_3$ metric can be recovered from the gauge fields
\begin{align}
g_{ij}=\frac{1}{2}\text{Tr}\Big[(A_{i}-\bar{A}_{i})(A_{j}-\bar{A}_{j})\Big].
\end{align}
\par
For the root-$T\bar{T}$-deformed BTZ black hole, which can be obtained from the BTZ black hole through a coordinate transformation. The AdS$_3$ gravity with mixed boundary condition can also described by two copies of SL$(2,\mathbb{R})$ Chern-Simons theory but with a non-trivial boundary term. In Chern-Simons formulation, the deformed gauge connection reads
\begin{align}
A(\lambda)&=-\frac{1}{r}L_0 dr+\left(-r\mathcal{L}_{\lambda} L_{-1}+\frac{1}{r}L_1\right)\left(\cosh\left(\frac{\lambda}{2}\right)dU-\sqrt{\frac{\overline{\mathcal{L}}_{\lambda}}{\mathcal{L}_{\lambda}}}\sinh\left(\frac{\lambda}{2}\right)dV\right),\\
\bar{A}(\lambda)&=\frac{1}{r}L_0 dr+\left(\frac{1}{r}L_{-1}-r\overline{\mathcal{L}}_{\lambda} 
L_1\right)\left(\cosh\left(\frac{\lambda}{2}\right)dV-\sqrt{\frac{\mathcal{L}_{\lambda}}{\overline{\mathcal{L}}_{\lambda}}}\sinh\left(\frac{\lambda}{2}\right)dU\right).
\end{align}
Moreover, since the deformed does not change the radial coordinate, the radial degree of freedom can also be eliminated through the gauge transformation
\begin{align}
A(\lambda)=b^{-1}\Big(d+a(\lambda)\Big)b,\quad \bar{A}(\lambda)=b\Big(d+\bar{a}(\lambda)\Big)b^{-1},\quad b=e^{\log r L_0},
\end{align}
where
\begin{align}
a(\lambda)&=\left(-\mathcal{L}_{\lambda} L_{-1}+L_1\right)\left(\cosh\left(\frac{\lambda}{2}\right)-\sqrt{\frac{\overline{\mathcal{L}}_{\lambda}}{\mathcal{L}_{\lambda}}}\sinh\left(\frac{\lambda}{2}\right)\right)d\theta\nonumber\\
&+\left(-\mathcal{L}_{\lambda} L_{-1}+L_1\right)\left(\cosh\left(\frac{\lambda}{2}\right)+\sqrt{\frac{\overline{\mathcal{L}}_{\lambda}}{\mathcal{L}_{\lambda}}}\sinh\left(\frac{\lambda}{2}\right)\right)d t,\\
\bar{a}(\lambda)&=\left(L_{-1}-\overline{\mathcal{L}}_{\lambda} 
L_1\right)\left(\cosh\left(\frac{\lambda}{2}\right)-\sqrt{\frac{\mathcal{L}_{\lambda}}{\overline{\mathcal{L}}_{\lambda}}}\sinh\left(\frac{\lambda}{2}\right)\right)d\theta\nonumber\\
&-\left(L_{-1}-\overline{\mathcal{L}}_{\lambda} 
L_1\right)\left(\cosh\left(\frac{\lambda}{2}\right)+\sqrt{\frac{\mathcal{L}_{\lambda}}{\overline{\mathcal{L}}_{\lambda}}}\sinh\left(\frac{\lambda}{2}\right)\right)dt.
\end{align}
In~\cite{Ebert:2023tih}, it was shown the boundary term associated with the root-$T\bar{T}$ deformed boundary condition is given by 
\begin{align}
S_{\mathrm{bdry}}=\frac{1}{2}\int_{\partial M}dtd\theta\left(\mathcal{L}_{\lambda}+\overline{\mathcal{L}}_{\lambda}\right)=\int_{\partial M}dtd\theta\sqrt{X_{\theta\theta}\overline{X}_{\theta\theta}}\sinh(\lambda)+\frac{1}{2}\cosh(\lambda)\left(X_{\theta\theta}+\overline{X}_{\theta\theta}\right)
\end{align}
where 
\begin{align}
X_{ij}=\text{Tr}\Big(A_i(\lambda)A_j(\lambda)\Big),\quad \bar{X}_{ij}=\text{Tr}\Big(\bar{A}_i(\lambda)\bar{A}_j(\lambda)\Big).
\end{align}
The boundary action exactly gives the root-$T\bar{T}$ deformed spectrum. With this setup we shall compute the root-$T\bar{T}$ deformed quantities from the AdS$_3$ gravity using the Chern-Simons formulation.
\subsection{Deformed BTZ black hole entropy}
In Chern-Simons formulation, the BTZ black hole entropy is given by the Wilson loop
\begin{align}
S_{\mathrm{th}}=-\log W_{\mathcal{R}}(\gamma_{\theta}),
\end{align}
and the Wilson loop is defined as
\begin{align}
W_{\mathcal{R}}(\gamma_{\theta})=\operatorname{tr}_{\mathcal{R}_f}\left(\mathcal{P}\exp\int_{\gamma_{\theta}}a\right)+\operatorname{tr}_{\mathcal{R}_f}\left(\mathcal{P}\exp\int_{\gamma_{\theta}}\bar{a}\right).
\end{align}
where $\mathcal{R}_f$ means the fundamental representation of $\mathrm{SL}(2,\mathbb{R})$ and the bilinear form is given by~\eqref{bilinear}. We first diagonalize the gauge connection
\begin{align}
a(\lambda)&=u \left(\sqrt{\mathcal{L}_{\lambda}}\cosh \left(\frac{\lambda }{2}\right)-\sqrt{\overline{\mathcal{L}}_{\lambda}}\sinh \left(\frac{\lambda }{2}\right)\right)L_0u^{-1},\\
\overline{a}(\lambda)&=\bar{u} \left(\sqrt{\overline{\mathcal{L}}_{\lambda}}\cosh \left(\frac{\lambda }{2}\right)-\sqrt{\mathcal{L}_{\lambda}}\sinh \left(\frac{\lambda }{2}\right)\right)L_0\bar{u}^{-1},
\end{align}
Then deformed Black hole entropy is given by
\begin{align}
S=2\pi\sqrt{\frac{c}{6}}\left(\sqrt{\mathcal{L}_{\lambda}}e^{-\frac{\lambda}{2}}+\sqrt{\overline{\mathcal{L}}_{\lambda}}e^{-\frac{\lambda}{2}}\right),\quad c=\frac{3l}{2G},
\end{align}
which is a Cardy-like formula. For the non-rotating BTZ black hole, we recover the result in field theory~\eqref{cardy_like}, which is obtained using the modular bootstrap.  
\subsection{One-loop partition functions} 
In this section, we consider the fields propagating in the root-$T\bar{T}$ deformed BTZ black hole background. The total action becomes
\begin{align}
S=-\frac{1}{16\pi G}\int_{\mathcal{M}} d^3 x\sqrt{g}(R+2)+\frac{1}{2}\int_{\mathcal{M}} d^3 x\sqrt{g}\mathcal{F}(-\Delta+m^2)\mathcal{F}+B,
\end{align} 
where $B$ is the boundary term. The fields $\mathcal{F}$ can be scalar fields, vector fields and gravitational fields, which are minimal coupling to the AdS$_3$ gravity. The partition function can be written into the path integral form 
\begin{align}
Z=\int[\mathcal{D}g_{\mu\nu}]e^{-I_{\mathrm{EH}}[g_{\mu\nu}]}Z_{\mathcal{F}}[g_{\mu\nu}],
\end{align}
where 
\begin{align}
Z_{\mathcal{F}}[g_{\mu\nu}]=\int[\mathcal{D}\phi]e^{-S_{\mathrm{m}}[g_{\mu\nu},\mathcal{F}]}.
\end{align}
\par
In the semi-classical limit $G\to 0$, only the Einstein-Hilbert action and the boundary term contribute to the partition function. Under the saddle point approximation, the partition function can be calculated by summing over all the gravitational saddles. The on-shell Euclidean action becomes the energy of AdS$_3$ times the euclidean time~\cite{Maldacena:1998bw}. For the root-$T\bar{T}$ deformed BTZ black hole, the leading semi-classical approximation to the partition function is
\begin{align}
Z_{\mathrm{classical}}=\exp\left(-\frac{\beta_2 (\mathcal{L}_{\lambda}+\overline{\mathcal{L}}_{\lambda})}{2}\right)=e^{-k e^{-\lambda}(\tau-\bar{\tau})},\quad k=\frac{1}{4G}.
\end{align} 
which is agree with the result in~\cite{Tian:2024vln}.
\par
The one-loop correction of partition function can be obtained by taking into account the contribution of  fields $\mathcal{F}$ propagating in the bulk on some gravitational saddles. Usually, the BTZ black holes and thermal AdS$_3$ are considered~\cite{Giombi:2008vd,David:2009xg}. The path integral for a fixed background can be calculated using the heat kernel method and the recently proposed Wilson spool method~\cite{Castro:2023bvo,Castro:2023dxp}. It turns out that these methods also work for the $T\bar{T}$-deformed AdS$_3$~\cite{He:2024pbp}. For the root-$T\bar{T}$ deformed BTZ black hole, we would like to use the Wilson spool proposal in terms of Chern-Simons formulation. 
\par
The partition function of massive scalar field in Chern-Simons AdS$_3$ gravity is described by the Wilson spool, which is a collection of Wilson loops winding around closed paths of the background~\cite{Castro:2023bvo,Castro:2023dxp}. In terms of the Chern-Simons formulation, the one-loop scalar field partition function in AdS$_3$ can be written as 
\begin{align}
Z^{\text{1-loop}}_{\mathrm{scalar}}[g_{\mu\nu}]=\exp\left(\frac{1}{4}\mathbb{W}_{j}[A,\bar{A}]\right),
\end{align}
where the Wilson spool is defined as
\begin{align}
\mathbb{W}_{j}[a,\bar{a}]=i\int_{\mathcal{C}}\frac{\mathrm{d}\alpha}{\alpha}\frac{\cos\alpha/2}{\sin\alpha/2}\mathrm{Tr}_{R_j}\left(\mathcal{P}e^{\frac{\alpha}{2\pi}\oint a}\right)\mathrm{Tr}_{R_j}\left(\mathcal{P}e^{-\frac{\alpha}{2\pi}\oint \bar{a}}\right).
\end{align}
Here $j$ labels a lowest-weight representation of $\mathrm{SL}(2,\mathbb{R})$ related to the mass of the bulk scalar field by
\begin{align}
\label{mass-j}
j=\frac{1}{2}\left(1+\sqrt{m^2+1}\right).
\end{align}
We have path ordered exponentials of $a$ and $\bar{a}$, which captures the information of the geometry. The contour integral is given by $\mathcal{C}=2\mathcal{C}_{+}$ with $\mathcal{C}_{+}$ running upwards along the imaginary $\alpha$ axis to the right of zero. This contour integral becomes a summing over poles implements a collection of Wilson loops with arbitrary windings. Then the Wilson spool gives the partition function of scalar field coupling with Chern-Simons theory. This proposal has been verified in AdS$_3$ and dS$_3$~\cite{Castro:2023bvo,Castro:2023dxp}. Recently, it turns out that the Wilson spool are tenable under $T\bar{T}$ deformation~\cite{He:2024pbp}.
\par
More generally, for the symmetric, transverse and traceless (STT) massive spin-$s$ field~\cite{Bourne:2024ded}, the partition function can also be calculated using the Wilson spool
\begin{align}
\mathbb{W}_{\Delta,s}[a,\bar{a}]&=i\sum_{\mathcal{R}_{\Delta,s}^{\mathrm{LW}}}\int_{\mathcal{C}_{+}}\frac{\mathrm{d}\alpha}{\alpha}\frac{\cos\left(\frac{\alpha}{2}\right)}{\sin\left(\frac{\alpha}{2}\right)}\mathrm{tr}_{\mathcal{R}_{j_L}}\left(\mathcal{P}e^{\frac{\alpha}{2\pi}\oint_{\gamma_{\mathrm{Sp}}}a}\right)\mathrm{tr}_{\mathcal{R}_{j_R}}\left(\mathcal{P}e^{-\frac{\alpha}{2\pi}\oint_{\gamma_{\mathrm{Sp}}}\bar{a}}\right)\nonumber\\
&+i\sum_{\mathcal{R}_{\Delta,s}^{\mathrm{HW}}}\int_{\mathcal{C}_{-}}\frac{\mathrm{d}\alpha}{\alpha}\frac{\cos\left(\frac{\alpha}{2}\right)}{\sin\left(\frac{\alpha}{2}\right)}\mathrm{tr}_{\mathcal{R}_{j_L}}\left(\mathcal{P}e^{\frac{\alpha}{2\pi}\oint_{\gamma_{\mathrm{Sp}}}a}\right)\mathrm{tr}_{\mathcal{R}_{j_R}}\left(\mathcal{P}e^{-\frac{\alpha}{2\pi}\oint_{\gamma_{\mathrm{Sp}}}\bar{a}}\right),
\end{align}
where
\begin{align}
\Delta=1+\sqrt{m^2+(s-1)^2},
\end{align}
which is satisfied for pairs of highest- and lowest-weight representations
\begin{align}
j_L=\frac{\Delta\pm s}{2},\quad j_R=\frac{\Delta\mp s}{2}.
\end{align}
We can also transform the Wilson spools into the lowest-weight representation
\begin{align}
\mathbb{W}_{j_{L},j_{R}}[a,\bar{a}]=i\sum_{\mathcal{R}_{\Delta,s}^{\mathrm{LW}}}\int_{2\mathcal{C}_{+}}\frac{\mathrm{d}\alpha}{\alpha}\frac{\cos\left(\frac{\alpha}{2}\right)}{\sin\left(\frac{\alpha}{2}\right)}\mathrm{tr}_{\mathcal{R}_{j_L}}\left(\mathcal{P}e^{\frac{\alpha}{2\pi}\oint_{\gamma_{\theta}}a}\right)\mathrm{tr}_{\mathcal{R}_{j_R}}\left(\mathcal{P}e^{-\frac{\alpha}{2\pi}\oint_{\gamma_{\theta}}\bar{a}}\right).
\end{align}
\par
We have learned that the root-$T\bar{T}$ deformed BTZ black hole is a quotient of hyperbolic space, and the deformed geometry is filled in $r$-$t$ plane. The $t$-cycle is bulk contractible while the $\theta$-cycle is not. We then choose the holonomies of the background connections $(a,\bar{a})$ around $\theta$-cycle. There are two main ingredients to calculate the Wilson spools. The one is  Wilson loop, which is the tracing over the different representation of the $\mathrm{SL}(2,\mathbb{R})$. The other one is the contour integral which can be calculated by summing all the residues. To calculate the Wilson loop in the root-$T\bar{T}$ deformed BTZ black hole, we first diagonalize the gauge fields
\begin{align}
\mathcal{P}\exp\left(\oint_{\gamma_{\theta}}a\right)&=u^{-1}e^{2\pi  hL_0}u,\\
\mathcal{P}\exp\left(\oint_{\gamma_{\theta}}\bar{a}\right)&=\bar{u}^{-1}e^{2\pi\bar{h}L_0}\bar{u},
\end{align}
where the $(h,\bar{h})$ can be written in terms of the modular parameters
\begin{align}
h=&2\left(\sqrt{\mathcal{L}_{\lambda}}\cosh \frac{\lambda }{2}-\sqrt{\overline{\mathcal{L}}_{\lambda}}\sinh \frac{\lambda }{2}\right)=-\frac{i}{2} \left(e^{-\lambda } (\tau -\bar{\tau})+\tau +\bar{\tau}\right),\\
\bar{h}=&2\left(\sqrt{\overline{\mathcal{L}}_{\lambda}}\cosh \frac{\lambda }{2}-\sqrt{\mathcal{L}_{\lambda}}\sinh \frac{\lambda }{2}\right)=\frac{i}{2} \left(e^{-\lambda } (\bar{\tau}-\tau)+\tau +\bar{\tau}\right).
\end{align}
The tracing over the lowest-weight representation are captured by the characters
\begin{align}
\chi_j(z)=\operatorname{Tr}_{R_j}\left(e^{2\pi zL_0}\right)=\frac{e^{\pi z(2j-1)}}{2\sinh(-\pi z)}.
\end{align}
After taking the residue of the contour integral at $\alpha=2n\pi, n\in \mathbb{Z}^{+}$, the partition function becomes
\begin{align}
\log(Z_{\Delta,\mathbf{s}})&=\frac{1}{4}\mathbb{W}_{j_{L},j_{R}}[a,\bar{a}]=\sum_{\pm}\sum_{n=1}^\infty \frac{1}{n}\chi_{j_L}(n h)\chi_{j_R}(-n\bar{h}).
\end{align}
We have to point out that the Wilson spool proposal depends on the DHS method~\cite{Denef:2009kn}, which is only hold for the STT massive spin-$s$ fields. The general vector fields and gravitation fields can be decomposed into the STT fields after we choose the proper gauge. Consequently, the partition functions for both vector fields and gravitational fields in AdS$_3$ and (root)-$T\bar{T}$ deformed  AdS$_3$ can be expressed in terms of STT fields, enabling their computation via Wilson spools.
\paragraph{Scalar field} For the scalar field, the Wilson spools can be straight calculated by taking the residues of the poles. We deform the contour to the right and pick the residues of the poles at $\alpha=2n\pi,n\in \mathbb{Z}^+$ for each $\mathcal{C}_{+}$. The result shows 
\begin{align}
\mathbb{W}_j[a,\bar{a}]=&\sum_{n=1}^\infty \frac{1}{n}\chi_j(n h)\chi_j(-n\bar{h})\nonumber\\
=&\sum_{n=1}^{\infty}\frac{e^{-i n \pi (2 j-1) e^{-\lambda } (\tau-\bar{\tau})}}{4 n \sin\left(\frac{n\pi}{2} \left(e^{-\lambda } (\tau -\bar{\tau})+\tau +\bar{\tau}\right)\right)\sin\left(\frac{n \pi}{2}\left(e^{-\lambda } (\bar{\tau}-\tau)+\tau +\bar{\tau}\right)\right)},
\end{align}
if we redefine the modular parameter
\begin{align}
\tau_{\lambda}=\frac{1}{2}\Big(\tau+\bar{\tau}+e^{-\lambda}(\tau-\bar{\tau})\Big),\quad \bar{\tau}_{\lambda}=\frac{1}{2}\Big(\tau+\bar{\tau}+e^{-\lambda}(\bar{\tau}-\tau)\Big).
\end{align}
The deformed modular parameter is also find in field theory~\eqref{new-modparameter}. Finally, the partition function can be written as 
\begin{align}
Z^{\text{1-loop}}_{\mathrm{scalar}}(\tau,\bar{\tau};\lambda)=\exp\left(\frac{1}{4}\mathbb{W}_{j}[A,\bar{A}]\right)=
\prod_{l,l'=0}^\infty\frac{1}{1-q_{\lambda}^{\ell+j}\bar{q}_{\lambda}^{\ell'+j}},
\end{align}
where
\begin{align}
\label{deformed_q}
q_{\lambda}=e^{2\pi i\tau_{\lambda}},\quad \bar{q}_{\lambda}=e^{-2\pi i\bar{\tau}_{\lambda}}.
\end{align}
This result agree with the result in field theory for the large $c$ limit.
\paragraph{Vector field}
For the vector field, the massive spin-$1$ fields can be decompose into STT spin-$1$ field and scalar field with the same mass~\cite{Christensen:1979iy,Denef:2009kn}. The partition function for massive field propagating in the deformed AdS$_3$ background can be written in terms of the Wilson spool
\begin{align}
Z_{\mathrm{vector}}^{\mathrm{1-loop}}(\tau,\bar{\tau};\lambda)=Z_{\Delta,1}(\tau,\bar{\tau};\lambda)\cdot Z_{\mathrm{scalar}}^{\mathrm{1-loop}}(\tau,\bar{\tau};\lambda),
\end{align}
where
\begin{align}
Z_{\Delta,1}(\tau,\bar{\tau};\lambda)&=\prod_{\ell, \ell' = 0}^{\infty} \frac{1}{(1 - q_{\lambda}^{\ell + j_L} \overline{q}_{\lambda}^{\ell'+j_{R}})(1 - q_{\lambda}^{\ell+j_{R}} \overline{q}_{\lambda}^{\ell' + j_{L}})}.
\end{align}
\paragraph{Gravitational field} For the gravitational fields, we consider the tensor perturbation of the root-$T\bar{T}$ deformed BTZ black hole. Since the root-$T\bar{T}$ deformed AdS$_3$ metric still satisfies the Einstein equation but with different boundary condition, we can use the same technique in AdS$_3$. We should consider the perturbation
\begin{align}
g_{\mu\nu}\to g_{\mu\nu}+h_{\mu\nu},
\end{align}
and find the action for the metric $h_{\mu\nu}$. Here, the perturbation $h_{\mu\nu}$ is the order of $O(\sqrt{G})$, so that it would contribute the one-loop correction. The action for $h_{\mu\nu}$ depends on the choice of gauge. One of the convenient choice of gauge in AdS gravity is introduced in~\cite{tHooft:1974toh, Christensen:1979iy}. The result shows the perturbative action can be separated into trace part and traceless part, which is described by the scalar field and the traceless symmetric tensor field, defined as
\begin{align}
\phi=h_{\rho}^{\rho},\quad \phi_{\mu\nu}=h_{\mu\nu}-\frac{1}{3}g_{\mu\nu}h^{\rho}_{\rho}.
\end{align}
In addition, the gauge-fixing procedure also introduces a Fadeev-Popov ghost field, which in this case is a complex valued vector field $\phi_{\mu}$. Including the graviton perturbation, we therefore have the action
\begin{align}
S=I_{\mathrm{EH}}[g_{\mu\nu}]+S_{\mathrm{trace}}[\phi]+S_{\mathrm{traceless}}[\phi_{\mu\nu}]+S_{\mathrm{ghost}}[\phi_{\mu}],
\end{align}
where
\begin{align}
S_{\mathrm{trace}}&=-\frac{1}{32\pi G}\int d^3x\sqrt{g}\left[\frac{1}{12}\phi\left(-\nabla^2+4\right)\phi\right],\nonumber\\
S_{\mathrm{traceless}}&=-\frac{1}{32\pi G}\int d^3x\sqrt{g}\left[\frac{1}{2}\phi_{\mu\nu}\left(g^{\mu\rho}g^{\nu\sigma}\nabla^2+2R^{\mu\rho\nu\sigma}\right)\phi_{\rho\sigma}
\right],\nonumber\\
S_{\mathrm{ghost}}&=\frac{1}{32\pi G}\int d^3x\sqrt{g}\phi^{*}_{\mu}\left(-g^{\mu\nu}\nabla^2-R^{\mu\nu}\right)\phi_{\nu}.
\end{align}
Then the one-loop partition function contains three parts
\begin{align}
\log Z_{\text{graviton}}^{\text{1-loop}}&=- \frac{1}{2}\log\det\Delta^{(2)}+\log\det\Delta^{(1)}-\frac{1}{2}\log\det\Delta^{(0)},
\end{align}
where
\begin{align}
\Delta^{(0)}=&-\nabla^2+4,\\
\Delta^{(1)}=&(-\nabla^2+4)g_{\mu\nu},\\
\Delta^{(2)}=&-\nabla^2g_{\mu\rho}g_{\mu\sigma}.
\end{align}
\par
The scalar field and ghost field parts have been discussed. The massless symmetric traceless rank-2 tensor can be decomposed into the massless STT spin-$2$ part and the massive vector part with $m^2=4$~\cite{Gibbons:1978ji}. Therefore both of them can be calculated using the Wilson spools. The partition function for massless spin-2 transverse part can be written as 
\begin{align}
-\frac{1}{2}\det\Delta^{(2)}=\log Z_{2,2}(\tau,\bar{\tau};\lambda)-\frac{1}{2}\det\Delta^{(1)},
\end{align}
where
\begin{align}
Z_{2,2}(\tau,\bar{\tau};\lambda)&=\prod_{\ell, \ell' = 0}^{\infty} \frac{1}{\left(1 - q_{\lambda}^{\ell} \overline{q}_{\lambda}^{\ell'+2}\right)\left(1 - q_{\lambda}^{\ell+2} \overline{q}_{\lambda}^{\ell' }\right)}.
\end{align}
The partition function for massive vector part with $m^2=4$ reads
\begin{align}
-\frac{1}{2}\det\Delta^{(1)}=\log Z_{3,1}(\tau,\bar{\tau})+\log Z_{\mathrm{scalar}}^{\mathrm{1-loop}}(\tau,\bar{\tau};\lambda),
\end{align}
where
\begin{align}
Z_{3,1}(\tau,\bar{\tau};\lambda)&=\prod_{\ell, \ell' = 0}^{\infty} \frac{1}{\left(1 - q_{\lambda}^{\ell + 1} \overline{q}_{\lambda}^{\ell'+2}\right)\left(1 - q_{\lambda}^{\ell+2} \overline{q}_{\lambda}^{\ell' + 1}\right)}.
\end{align}
The partition function for massive scalar part with $m^2=4$ reads
\begin{align}
-\frac{1}{2}\log\det\Delta^{(0)}=\log Z_{\mathrm{scalar}}^{\mathrm{1-loop}}(\tau,\bar{\tau};\lambda),
\end{align}
where
\begin{align}
Z_{\mathrm{scalar}}^{\mathrm{1-loop}}(\tau,\bar{\tau};\lambda)&=\prod_{\ell, \ell' = 0}^{\infty} \frac{1}{\left(1 - q_{\lambda}^{\ell+\frac{\sqrt{5}+1}{2}} \overline{q}_{\lambda}^{\ell'+\frac{\sqrt{5}+1}{2}}\right)}.
\end{align}
\par
Finally, combining all these parts, we obtain the one-loop partition function for the graviton
\begin{align}
Z^{\mathrm{1-loop}}_{\mathrm{graviton}}=&\frac{Z_{3,1}(\tau,\bar{\tau};\lambda)}{Z_{2,2}(\tau,\bar{\tau};\lambda)}\nonumber\\
=&\exp\left(\sum_{n=1}^{\infty}\frac{q_{\lambda}^{2n}+\bar{q}_{\lambda}^{2n}}{n|1-q_{\lambda}^n|^2}-\sum_{n=1}^{\infty}\frac{|q_{\lambda}|^{2n}(q_{\lambda}^n+\bar{q}_{\lambda}^n)}{n|1-q_{\lambda}^n|^2}\right)\nonumber\\
=&\prod_{n=2}^{\infty}\frac{1}{|1-q_{\lambda}^n|^2},
\end{align}
Taking into account the tree level contribution, the full gravity partition function is
\begin{align}
Z_{\mathrm{gravity}}=|q_{\lambda}|^{-2k}\prod_{n=2}^{\infty}\frac{1}{|1-q_{\lambda}^n|^2}.
\end{align}
This result agrees with the root-$T\bar{T}$ deformed characters~\eqref{rootttbar_character}. This formula has a natural physical interpretation, which is the tracing over an irreducible representation of the Virasoro algebra
\begin{align}
Z_{\mathrm{gravity}}=\mathrm{Tr}\left(q_{\lambda}^{L_0-\frac{c}{24}}\bar{q}_{\lambda}^{\bar{L}_0-\frac{c}{24}}\right).
\end{align}
This representation contains a ground state of weight $k$, along with its Virasoro descendants. For the undeformed case, the result comes from the observation of that the asymptotic symmetry group of AdS$_3$ with Brown-Henneaux boundary condition is the Virasoro algebra~\cite{Brown:1986nw}. The similar result was found for $T\bar{T}$ deformation, which was interpreted as that the asymptotic symmetry group for AdS$_3$ with mixed boundary condition is a state-dependent Virasoro algebra~\cite{Guica:2019nzm, He:2021bhj}. From the one-loop gravity partition function, we infer that the asymptotic symmetry group of AdS$_3$ with the root-$T\bar{T}$ deformed boundary condition may be also described by the state-dependent Virasoro algebra. In contrast to AdS$_3$ with Brown-Henneaux boundary conditions, the deformed partition function is not one-loop exact and receives higher order corrections in this case. 
\section{Conclusion and discussion}
\label{sec:5}
The root-$T\bar{T}$ deformed CFT is shown dual to the AdS$_3$ with certain modified boundary condition. In this paper, we consider the root-$T\bar{T}$ deformed CFT partition function form both field theory side and gravity side. On the field theory side, we first analyse the zero-momentum case and demonstrate that the deformed partition function corresponds to a redefinition of the modular parameter. We show that the deformed partition function satisfies a flow equation and remains modular invariant for the redefined modular parameters. We then generalize these results to the case of arbitrary momentum. Through the perturbative expansion of the partition function, we derive a flow equation. In the large $c$ limit, the deformed partition function reduces to that of the original CFT, up to a redefinition of the modular parameter. Crucially, the partition function retains invariance under the modular transformations of redefined modular parameters. Using the modular bootstrap trick, we show that the asymptotic density of states is governed by a Cardy-like formula.
\par 
On the gravity side, we first show the root-$T\bar{T}$ deformed BTZ black hole is a quotient of hyperbolic space. We then reformulate the AdS$_3$  with root-$T\bar{T}$ deformed boundary conditions in the Chern-Simons framework. Within this Chern-Simons formulation, we compute both the deformed BTZ black hole entropy and the one-loop partition function of the deformed AdS$_3$ using the Wilson loops and Wilson spools techniques, respectively. Our results reveal that the entropy of the deformed BTZ black hole follows a Cardy-like formula, in precise agreement with the field theory results derived via modular bootstrap trick. Furthermore, the one-loop partition function exactly matches the large $c$ expansion of the root-$T\bar{T}$ deformed CFT partition function. These results provide the evidence for the proposed duality between root-$T\bar{T}$ deformed CFTs and AdS$_3$ gravity with root-$T\bar{T}$ deformed boundary condition.
\par
The one-loop partition function is described by tracing over a state-dependent Virasoro algebra, analogous to the case of $T\bar{T}$ deformation. This suggests that the asymptotic symmetry of AdS$_3$ with root-$T\bar{T}$ deformed boundary condition corresponds to a state-dependent Virasoro algebra. A natural extension would be to investigate the asymptotic symmetry algebra for AdS$_3$ under root-$T\bar{T}$ deformation. Furthermore, we observe that the root-$T\bar{T}$ deformed partition function loses modular invariance beyond the large $c$ limit, as evidenced by its flow equation. However, an interesting exception arises when the initial partition function is given by a $(1,1)$ weight modular form, the flow equation preserves the modular properties in this case. Notably, the seed theory is no longer the conformal field theory. 
\par
There are also some interesting topics for further study. In this work we only consider the large $c$ limit of the deformed partition function, up to $c^0$ order. The complete partition function is not modular invariant, although it obeys the flow equation. An important open question is how to systematically characterize the modular properties of root-$T\bar{T}$ deformed CFT partition functions beyond the large $c$ limit. In addition, it is shown that the root-$T\bar{T}$ deformations can be classically formulated by coupling the undeformed theory with 2D gravity action in~\cite{Babaei-Aghbolagh:2024hti}. It would be interesting to derive the partition function flow equation from the 2D gravity following the case for $T\bar{T}$ deformation~\cite{Dubovsky:2018bmo}. Moreover, one can also use the perturbation expansion of the partition function to consider the higher order corrections and study the resurgence of root-$T\bar{T}$ deformed CFT partition function following the recent papers for $T\bar{T}$ deformation~\cite{Gu:2024ogh,Gu:2025tpy}. 
\par
Finally, another interesting topic is to study the integrability of root-$T\bar{T}$-deformation on quantum level. The $T\bar{T}$ deformation manifests its effect by introducing the CDD factor multiplication into the S-matrix. We can obtain the spectrum and partition function using the Thermodynamic Bethe ansatz. Whether can we find the similar properties for the root-$T\bar{T}$ deformation. We may also have the form factor bootstrap program to calculate the correlation functions and entanglement entropy under the root-$T\bar{T}$ deformation following~\cite{Castro-Alvaredo:2023rtl,Castro-Alvaredo:2023wmw,Castro-Alvaredo:2023jbg,He:2023obo}, as well as the root-$T\bar{T}$ deformation with boundary or defects~\cite{Jiang:2021jbg,Brizio:2024doe,Castro-Alvaredo:2025nma}.
\section*{Acknowledgments}
I am grateful to Yunfeng Jiang for helpful discussions and comments on this manuscript. I also thank Jie Gu, Yi-Jun He, Jue Hou, Yang Lei and Yuan Sun for helpful discussions. 
\appendix
\section{Basics on modular form}
\label{app:1}
In this section, we discuss the modular properties for the deformed partition function based on the flow equation. We first give some basics about the modular form. We then show the root-$T\bar{T}$ deformed partition function have no obvious modular property
\par
The modular function with weight $(k,k')$ is a function transformed as 
\begin{align}
f_{k,k'}\left(\gamma\cdot \tau,\gamma\cdot \bar{\tau}\right)=(c\tau +d)^k (c\bar{\tau} +d)^{k'}f_{k,k'}(\tau,\bar{\tau}),
\end{align}
where
\begin{align}
\gamma\cdot \tau=\frac{a\tau+b}{c\tau+d},\quad \gamma\cdot \bar{\tau}=\frac{a\bar{\tau}+b}{c\bar{\tau}+d},\quad ad-bc=1\quad  a,b,c,d\in \mathbb{Z}.
\end{align}
The function is  modular invariant for $k=k'=0$. It is useful to understand the action of the derivatives $\partial_{\tau}$ and $\partial_{\bar{\tau}}$ on modular functions of various weights. By using the chain rule, one can easily find that the derivative $\partial_{\tau}$ and $\partial_{\bar{\tau}}$ transforms under modular transformation as
\begin{align}
\partial_{\tau}=\partial_{\tau}(\gamma\cdot\tau)\partial_{\gamma\cdot\tau}=\frac{1}{(c\tau+d)^{2}}\partial_{\gamma\cdot\tau},\\
\partial_{\bar{\tau}}=\partial_{\tau}(\gamma\cdot\bar{\tau})\partial_{\gamma\cdot\bar{\tau}}=\frac{1}{(c\bar{\tau}+d)^{2}}\partial_{\gamma\cdot\bar{\tau}}.
\end{align}
which means the derivative $\partial_{\tau}$ and $\partial_{\bar{\tau}}$ act on a modular invariance function ($(0,0)$ modular form) ending up with the (2,0) and (0,2) weight modular form, respectively.
We also have
\begin{align}
\partial_{\tau}\partial_{\bar{\tau}}=&\frac{1}{(c\tau+d)^{2}}\frac{1}{(c\bar{\tau}+d)^{2}}\partial_{\gamma\cdot\tau}\partial_{\gamma\cdot\bar{\tau}},
\end{align}
which act on the modular invariant function ending up with a $(2,2)$ modular form. More generally, we have the modular covariant derivatives 
\begin{align}
\mathsf{D}_\tau^{(k)}=\partial_\tau-\frac{ik}{2\tau_2},\quad\mathsf{D}_{\bar{\tau}}^{(k')}=\partial_{\bar{\tau}}+\frac{ik'}{2\tau_2}.
\end{align}
This is also called the Ramanujan-Serre derivative. $\mathsf{D}_\tau^{(k)}$ acts on a modular form of weight $(k,k')$ gives a modular form of weight $(k+2,k')$. Similarly, $\mathsf{D}_{\bar\tau}^{(k')}$
increases the weight of such a modular form to $(k,k'+2)$. Generally, we have
\begin{align}
\mathsf{D}_\tau^{(k)}\mathsf{D}_{\bar{\tau}}^{(k')}=\partial_{\tau}\partial_{\bar{\tau}}+\left(\frac{ik'}{2\tau_2}\partial_{\tau}-\frac{ik}{2\tau_2}\partial_{\bar{\tau}}\right)+\frac{k'(k-1)}{4\tau_2^2}
\end{align}
which map the $(k,k')$ weight modular form into a $(k+2,k'+2)$ weight modular form. The simplest one is
\begin{align}
\mathscr{D}^{(0)}\equiv\mathsf{D}_\tau^{(0)}\mathsf{D}_{\bar{\tau}}^{(0)}=\partial_\tau\partial_{\bar{\tau}},
\end{align} 
also we introduce the following notation
\begin{align}
\mathscr{D}^{(k)}\equiv\mathsf{D}_\tau^{(k)}\mathsf{D}_{\bar{\tau}}^{(k)}=\partial_{\tau}\partial_{\bar{\tau}}+\frac{k}{2\tau_2}\partial_{\tau_2}+\frac{k(k-1)}{4\tau_2^2}.
\end{align}
In addition, it should be note that $\tau_2=\mathrm{Im}\tau$ is a modular form of weight $(-1,-1)$. 
\par
The root-$T\bar{T}$ deformed partition function has the following expansion
\begin{align}
\mathcal{Z}(\tau,\bar{\tau}|\lambda)=\sum_{n=0}^\infty\lambda^{n}Z_n(\tau,\bar{\tau}) .
\end{align}
In section~\ref{sec:3.2}, we have shown the perturbation expansion can be written in terms of some derivative operators act on the original partition function
\begin{align}
Z_p=
\left\{
  \renewcommand{\arraystretch}{2}  
     \begin{array}{lr}  
\left(\sum_{m=1}^{p}\tau_2^m\widehat{\mathcal{O}}_m^{(p)}\right)Z_0(\tau,\bar{\tau}),&\text{for even $p$}\\
\left(\sum_{m=1}^{p}\tau_2^m\widehat{\mathcal{O}}_m^{(p)}\right)\widehat{\mathcal{R}}Z_0(\tau,\bar{\tau}),&\text{for odd $p$}
\end{array}
\right.
\end{align}
where
\begin{align}
\widehat{\mathcal{R}}Z_0=\sum_{n=0}^{\infty}2 \pi R \sqrt{E_n^2-P_n^2}e^{2\pi i R \tau_1P_n-2\pi R\tau_2E_n}.
\end{align}
We find 
\begin{align}
\frac{1}{(2\pi R)^2}\left(\partial_{\tau_1}^2+\partial_{\tau_2}^2\right)e^{2\pi i R \tau_1P_n-2\pi R\tau_2E_n}=\left(E_n^2-P_n^2\right)e^{2\pi i R \tau_1P_n-2\pi R\tau_2E_n},
\end{align}
which means $e^{2\pi i R \tau_1P_n-2\pi R\tau_2E_n}$ is the eigenstate of the operator $\left(\partial_{\tau_1}^2+\partial_{\tau_2}^2\right)$. Formally, we can define the square-root operator
\begin{align}
\widehat{\mathcal{R}}\equiv\sqrt{\partial_{\tau_1}^2+\partial_{\tau_2}^2}=2\sqrt{\partial_{\tau}\partial_{\bar{\tau}}},
\end{align}
which act on the $e^{2\pi i R \tau_1P_n-2\pi R\tau_2E_n}$ with a square root of the eigenvalue
\begin{align}
\widehat{\mathcal{R}}\cdot e^{2\pi i R \tau_1P_n-2\pi R\tau_2E_n}=2\pi R\sqrt{E_n^2-P_n^2}e^{2\pi i R \tau_1P_n-2\pi R\tau_2E_n}.
\end{align}
Therefore, we have the following replacements
\begin{align}
2\pi R\sqrt{E_n^2-P_n^2}\mapsto \widehat{\mathcal{R}}\equiv\sqrt{\partial_{\tau_1}^2+\partial_{\tau_2}^2}=2\sqrt{\partial_{\tau}\partial_{\bar{\tau}}}.
\end{align}
In fact, the power of a generic operator is defined as
\begin{align}
(\mathcal{O})^\alpha=\frac{1}{\Gamma(-\alpha)}\int_0^\infty ds s^{-\alpha-1}\mathrm{e}^{-s\mathcal{O}},
\end{align}
which is used to calculate root-$T\bar{T}$ deformed correlation function~\cite{Hadasz:2024pew}. The square-root operator can also be written as the Ramanujan-Serre derivative
\begin{align}
\widehat{\mathcal{R}}=2\sqrt{\partial_{\tau}\partial_{\bar{\tau}}}=2\sqrt{\mathscr{D}^{(0)}}.
\end{align}
In order to avoid the square root of the differential operator, we consider the second order recurrence relation and the result reads
\begin{align}
Z_{p+2}=\frac{1}{(p+1)(p+2)}\Big(4\tau_2^2\left(\partial_{\tau}\partial_{\bar{\tau}}\right)+\tau_2\partial_{\tau_2}\Big)Z_p,
\end{align} 
which holds for both even and odd orders. The recurrence relation implies the flow equation
\begin{align}
\partial_{\lambda}^{2}\mathcal{Z}(\tau,\bar{\tau}|\lambda)=\left(4\tau_{2}^{2}\left(\partial_{\tau}\partial_{{\bar{\tau}}}\right)+\tau_{2}\partial_{{\tau_{2}}}\right)\mathcal{Z}(\tau,\bar{\tau}|\lambda),
\end{align}
with the initial condition
\begin{align}
\mathcal{Z}(\tau,\bar{\tau}|\lambda)\Big|_{\lambda=0}=Z_0,\quad \partial_{\lambda}\mathcal{Z}(\tau,\bar{\tau}|\lambda)\Big|_{\lambda=0}=-2\tau_2\widehat{\mathcal{R}}Z_0.
\end{align}
In the case of $T\bar{T}$ deformation the recurrence relation is given by a Ramanujan-Serre derivative, which has certain modular properties. However, we do not find the modular properties from the flow equation for the root-$T\bar{T}$ deformation. To see this, we first rewrite the recurrence relation in terms of Ramanujan-Serre derivative
\begin{align}
Z_{p+2}=\frac{4}{(p+1)(p+2)}\tau_2^2\mathscr{D}^{(1)}Z_{p}.
\end{align}
This is second order recurrence relation, we have to give the initial condition
\begin{align}
Z_0\quad\text{and}\quad Z_1=-2\tau_2\sqrt{\mathscr{D}^{(0)}}Z_0.
\end{align}
If we start from a modular invariant partition function $Z_0$, the first order correction $Z_1$ is modular invariant since the $\mathscr{D}^{(0)}$ map (0,0) modular form into (2,2) modular form and $\tau_2$ is a $(-1,-1)$ form. While the higher order correction is not obviously modular invariant because $\mathscr{D}^{(1)}$ acts on modular invariant object will end up with a complicated result, which may not be a certain modular form in general. However, We can not rule out the root-$T\bar{T}$ deformed partition function is modular invariant either.
\section{Another expansion for the partition function}
\label{app:2}
In this appendix, we show another expansion of the root-$T\bar{T}$ deformed partition function. In generally, we do the perturbative expansion for the deformation parameter. We find another expansion which is a Laurent expansion of parameter $t=e^{-\lambda}$. This expansion also provide a check of that the deformed partition function exactly satisfy the flow equation~\eqref{partition_flow_equation}.
\par
We start from the root-$T\bar{T}$ deformed partition function, which can be written as 
\begin{align}
\mathcal{Z}(\tau,\bar{\tau}|\lambda)=&\sum_{n=0}^{\infty}e^{-2 \pi  R \tau_2 \cosh(\lambda) E_n-2\pi R \tau_2\sinh(\lambda)\sqrt{E_n^2-P_n^2}+2 \pi  i R \tau_1 P_n}\nonumber\\
=&\sum_{n=0}^{\infty}\exp\left(\pi R\tau_2 \left(A_nt-\frac{B_n}{t}\right)+2 \pi R \tau_1\sqrt{A_nB_n}\right),
\end{align}
where we have defined 
\begin{align}
t=e^{-\lambda},\quad A_n=\sqrt{E_n^2-P_n^2}-E_n,\quad B_n=\sqrt{E_n^2-P_n^2}+E_n.
\end{align}
Note the identity
\begin{align}
\exp{\left(\frac{A_n t x}{2}-\frac{B_n x}{2 t}\right)}=\sum _{k=-\infty }^{\infty }J_k\left(x \sqrt{A_n B_n}\right)\left(\sqrt{\frac{A_n}{B_n}}\right)^kt^k,
\end{align}
which provide an Laurent expansion of parameter $t$.
So that the partition function have the following expansion
\begin{align}
\mathcal{Z}(\tau,\bar{\tau}|t)=\sum _{k=-\infty }^{\infty }\sum_{n=0}^{\infty}e^{2\pi R\tau_1\sqrt{A_n B_n}}J_k\left(2\pi R\tau_2\sqrt{A_n B_n}\right)\left(\sqrt{\frac{A_n}{B_n}}\right)^kt^k,
\end{align}
where the $J$-Bessel function is given by
\begin{align}
J_\nu(x) = \sum_{k=0}^{\infty} \frac{(-1)^k}{k! \Gamma(k+\nu+1)} \left( \frac{x}{2} \right)^{2k+\nu}.
\end{align}
One can verify it satisfies the flow equation~\eqref{partition_flow_equation}. In addition, for $t=1$, we have
\begin{align}
\mathcal{Z}(\tau,\bar{\tau}|1)=&\sum _{k=-\infty }^{\infty }\sum_{n=0}^{\infty}e^{2\pi R\tau_1\sqrt{A_n B_n}}J_k\left(2\pi R\tau_2\sqrt{A_n B_n}\right)\left(\sqrt{\frac{A_n}{B_n}}\right)^k\nonumber\\
=&\sum_{n=0}^{\infty}e^{2\pi i R \tau_1P_n-2\pi R\tau_2E_n}.
\end{align}
\bibliographystyle{utphys.bst}
\bibliography{reference}
\end{document}